%% file: main.tex
\PassOptionsToPackage{table,xcdraw,dvipsnames}{xcolor}
\documentclass[sigconf,screen]{acmart}

\usepackage{listings}
\usepackage{enumitem}
\usepackage{tcolorbox}
\tcbuselibrary{listings,skins,breakable}
\usepackage[normalem]{ulem}

\usepackage{caption}
\AtBeginDocument{%
  }

\lstdefinelanguage{Python}{
    morekeywords={class, def, return, try, except, raise, from},
    keywordstyle=\color{blue},
    stringstyle=\color{green},
    commentstyle=\color{gray},
    morecomment=[l]{\#},
}

\lstdefinelanguage{diff}{
    morecomment=[f][\color{green}]{+},
    morecomment=[f][\color{red}]{-},
    morecomment=[f][\color{blue}]{@@},
}

\lstdefinelanguage{errorlog}{
    morecomment=[f][\color{red}]{E},
    morecomment=[f][\color{magenta}]{?},
}

\usepackage{algorithm}
\usepackage{algorithmic}

\usepackage{listings}
\usepackage{xspace}

\usepackage[T1]{fontenc}

\usepackage[utf8]{inputenc}

\usepackage{microtype}

\usepackage{graphicx}

\usepackage{subcaption}
\usepackage{listings}
\usepackage{upquote}

\usepackage{soul}
\usepackage{framed}
\usepackage{multirow}
\usepackage{siunitx}

\newcommand*\colourcheck[1]{%
	\expandafter\newcommand\csname #1check\endcsname{\textcolor{#1}{\ding{52}}}%
}
\colourcheck{blue}
\colourcheck{green}
\colourcheck{red}

\newcommand{\tool}{\textsc{CodeSteer}\xspace}
\newcommand{\tracecue}[1]{{\color{RoyalBlue}#1}}
\newcommand{\tracecf}[1]{{\color{BurntOrange}#1}}
\newcommand{\tracestate}[1]{{\color{ForestGreen!70!black}#1}}
\newcommand{\tracefinal}[1]{{\color{Purple}#1}}

\newcommand{\examplebanner}[2]{%
  \par\medskip
  \noindent\colorbox{#1!12}{\parbox{\dimexpr\linewidth-2\fboxsep\relax}{\textbf{#2}}}%
  \par\smallskip
}

\lstdefinestyle{motivationjava}{
  language=Java,
  basicstyle=\ttfamily\scriptsize,
  keywordstyle=\bfseries,
  columns=fullflexible,
  keepspaces=true,
  breaklines=true,
  showstringspaces=false,
  frame=single,
  framerule=0.35pt,
  rulecolor=\color{black!35},
  xleftmargin= 2.5em,
  xrightmargin= 1em,
  aboveskip=0.35em,
  belowskip=0.35em
}

\newboolean{showcomments}
\setboolean{showcomments}{true}
\ifthenelse{\boolean{showcomments}}
 { \newcommand{\mynote}[2]{
      \fbox{\bfseries\sffamily\scriptsize#1}
        {\small$\blacktriangleright$\textsf{\emph{#2}}$\blacktriangleleft$}}}
        { \newcommand{\mynote}[2]{}}

\newcommand{\cmark}{\ding{51}}%

\newcolumntype{L}[1]{>{\raggedright\arraybackslash}p{#1}}

\newcommand{\code}[1]{{\footnotesize\texttt{#1}}}
\usepackage{amsthm}
\definecolor{dkgreen}{rgb}{0,0.45,0}
\definecolor{gray}{rgb}{0.5,0.5,0.5}
\definecolor{lightgray}{rgb}{211, 211, 211}
\definecolor{mauve}{rgb}{0.58,0,0.82}

\definecolor{custom-red}{rgb}{0,0,0}
\definecolor{custom-blue}{rgb}{0,0,0}

\definecolor{c1}{HTML}{f4cccc}
\definecolor{c2}{HTML}{f5cdcd}
\definecolor{c3}{HTML}{fffcfc}
\definecolor{c4}{HTML}{ffffff}
\definecolor{c5}{HTML}{ffffff}
\definecolor{c6}{HTML}{fffdfd}
\definecolor{c7}{HTML}{f5cfcf}
\definecolor{c8}{HTML}{fffbfb}
\definecolor{c9}{HTML}{ffffff}
\definecolor{c10}{HTML}{fffdfd}
\definecolor{c11}{HTML}{fefafa}
\definecolor{c12}{HTML}{fef7f7}
\definecolor{c13}{HTML}{ffffff}
\definecolor{c14}{HTML}{fffefe}
\definecolor{c15}{HTML}{ffffff}
\definecolor{c16}{HTML}{fefafa}
\definecolor{c17}{HTML}{fdf3f3}
\definecolor{c18}{HTML}{fffefe}
\definecolor{c19}{HTML}{fdf5f5}
\definecolor{c20}{HTML}{ffffff}

\setcopyright{cc}
\setcctype{by-nc-nd}
\acmDOI{10.1145/3832783.3837556}
\acmYear{2026}
\copyrightyear{2026}
\acmISBN{979-8-4007-2882-2/2026/10}
\acmConference[ASE '26]{Proceedings of the 41st IEEE/ACM International Conference on Automated Software Engineering}{October 12--16, 2026}{Munich, Germany}
\acmBooktitle{Proceedings of the 41st IEEE/ACM International Conference on Automated Software Engineering (ASE '26), October 12--16, 2026, Munich, Germany}
\acmSubmissionID{ase26main-p3738-p}
\received{2026-03-26}
\received[accepted]{2026-06-18}

\begin{document}



\title{Post-hoc Attention Steering of Large Language Models for Robust Code Understanding under Obfuscation}





\author{Xiaokai Rong}
\orcid{0009-0000-8457-8528}
\affiliation{%
  \institution{University of Texas at Dallas}
  \city{Dallas}
  \country{USA}
}
\email{xiaokai.rong@utdallas.edu}

\author{Aashish Yadavally}
\orcid{0000-0001-8785-6319}
\affiliation{%
  \institution{University of Central Florida}
  \city{Orlando}
  \country{USA}
}
\email{aashish.yadavally@ucf.edu}

\author{Tien N. Nguyen}
\correspondingauthor
\orcid{0009-0006-7962-6090}
\affiliation{%
  \institution{University of Texas at Dallas}
  \city{Dallas}
  \country{USA}
}
\email{tien.n.nguyen@utdallas.edu}













\begin{abstract}
Code obfuscation is widely used in software systems and malware to conceal program logic and hinder analysis, posing significant challenges for both human developers and automated tools. While large language models (LLMs) have shown strong capabilities in code understanding, 
our preliminary study shows that LLM performance significantly degrades on obfuscated code, suggesting a reliance on superficial lexical cues rather than deep semantic reasoning. To address this limitation, we propose {\tool}, a novel \emph{attention steering} approach that reallocates model attention toward semantically relevant program elements, including backward slices for output prediction and control-flow paths for execution reasoning. Our method integrates lightweight program analysis with inference-time attention steering to guide LLMs toward the core input-to-output dependencies of a program. Experiments across multiple models and datasets demonstrate that {\tool} significantly improves performance on obfuscated code, often recovering comparable accuracy to the level of unobfuscated programs. We also show {\tool}'s practical utility through a case study on buffer overflow detection, highlighting its potential for malware/vulnerability analysis and reverse engineering of obfuscated~code.

\end{abstract}






\maketitle

\input{sections/intro}

\input{sections/motivation-example}
\input{sections/preliminary-study}


\input{sections/technical}

\input{sections/emperical}
\input{sections/RQ1-new}
\input{sections/Stratified_Result_by_Obf_type}
\input{sections/Behavioral-analysis}

\input{sections/RQ4-new}

\input{sections/case-study}

\input{sections/related}

\vspace{-3pt}
\section{Threats to Validity and Conclusion}


\emph{Internal validity.}
Threats include errors in the obfuscation pipeline, slice construction, and steering implementation. 
Results may depend on prompts, calibration choices, and hyperparameters. 

\emph{External validity.}
Our study is limited to a small set of LLMs, and benchmark tasks focused on output and execution prediction. The findings may not fully generalize to other languages, obfuscation types, security workflows, despite the promising case study in C.

\emph{Construct validity.}
Our tasks do not cover all aspects of program comprehension. Our slice is an approximation of semantic relevance rather than a fully sound one, so the results are viewed as evidence of improved task reasoning rather than full code understanding.

{\tool} combines program analysis with attention steering to recover the performance that LLMs lose under obfuscation in various tasks. Our behavioral analysis suggests that steering helps shift the model away from System 1's lexical shortcuts and toward more System 2, code-grounded stepwise reasoning under~obfuscation. 


 
\section{Data Availability Statement}

All code and data is available at~\cite{codesteer2026}.

\newpage

\balance

\bibliographystyle{ACM-Reference-Format}

\bibliography{references,apr-references}


\end{document}

%% file: sections/intro.tex
\vspace{-1pt}
\section{Introduction}
\label{sec:intro}

Malware frequently leverages code obfuscation to disguise its attack mechanisms and internal logic, posing challenges for program analysis and reverse engineering. Code obfuscation is also widely used in software to protect intellectual property and conceal business logic. Obfuscation techniques transform a program into a semantically equivalent form that is intentionally difficult for humans to read or understand. Typical transformations include identifier renaming, control-flow flattening, dead-code insertion, call indirection, etc. While these transformations preserve code functionality, they significantly reduce readability and increase the cognitive effort required for understanding program behavior \cite{collberg1997taxonomy,collberg2002manufacturing, Obfuscation-The-Hidden-Malware, An-Observational-Investigation-of-Reverse-Engineers-Process-and-Mental-Models}.

Prior research has shown the substantial impact of obfuscation on human comprehension. Nguyen {\em et al.}~\cite{nguyen2026effectcodeobfuscationhuman} reported that~obfuscated code significantly reduces humans' accuracy in predicting program outputs. Their study further revealed an interesting cognitive phenomenon: when variable names are adversarially renamed to misleading identifiers, developers shift from fast, memory-based reasoning (System 1) to slower analytical reasoning (System 2)~\cite{nguyen2026effectcodeobfuscationhuman}. Meaningful identifier names, thus, serve as cognitive shortcuts for code comprehension, and removing those cues forces humans to rely on stepwise reasoning over program flows.

Large language models (LLMs) have demonstrated impressive capabilities in tasks such as code generation, summarization, vulnerability detection, and program reasoning \cite{chen2021evaluatinglargelanguagemodels,li2022codebert,guo2021graphcodebert,roziere2023codegen, Competition-level-code-generation-with-AlphaCode}. Most evaluations of these models are conducted on naturally written~source code where identifier names, comments, and common coding patterns provide strong semantic signals. With the rapid emergence of LLMs for code understanding and generation, an important open question arises: \textit{whether LLMs can maintain their reasoning capabilities when these signals are intentionally disrupted through obfuscation?} Answering this question has important implications for software security and analysis. If LLMs remain robust to obfuscation, they could become powerful tools for {\em analyzing and reverse engineering malware}. Conversely, if LLMs are confused by obfuscated code in a manner similar to humans, the effectiveness of~current obfuscation techniques may extend to AI-based~analysis. 


Our preliminary investigation reveals that LLMs are indeed significantly affected by obfuscation. Across several models, the accuracy of program output prediction decreases substantially when programs are obfuscated, even though the semantics of the code remain unchanged. This is consistent with the findings by Nikiema {\em et al.}~\cite{nikiema2025codebarrierllmsactually} as they reported a statistically significant performance decline as obfuscation complexity increases. This observation suggests that modern LLMs rely heavily on superficial lexical cues such as identifier names and familiar coding patterns rather than purely reasoning on program semantics.
This is particularly concerning in security-critical settings, e.g., {\em malware analysis}.
The inability of LLMs to robustly reason about obfuscated code may therefore limit their applicability in software security. Addressing this gap is crucial for {\em enabling reliable AI-assisted analysis of adversarial~code}.

Improving LLMs' code understanding on obfuscated code requires mechanisms that guide the models to focus on the program elements that are truly relevant for the task at hand, e.g., program output and execution prediction, rather than the above superficial lexical cues. A key challenge is that modern LLMs process code primarily as token sequences, and their attention mechanisms may distribute focus broadly across tokens that contain superficial lexical cues rather than the important statements that mainly contribute to decide the program's behavior. When those cues are removed or misleading as in obfuscated code, the model may fail to concentrate on crucial semantic dependencies.

{\color{custom-blue}
This motivates us to propose {\tool}, which explores {\bf attention reallocation} (or steering) as a training-free vehicle to guide LLM reasoning. Attention steering has emerged as a practical mechanism for influencing the reasoning trajectory of transformer models by biasing attention toward important tokens~\cite{PASTA, auto-pasta-zhang2024modeltellsattendfaithfulness}. 
{\tool} uses lightweight static program analysis to construct a soft relevance prior over code tokens including static slicing, data/control dependencies, and calling relations. This prior helps a model to pay attention to important tokens in the absence of lexical cues. 
To achieve that, {\tool} aims to steer LLM attention toward these semantically critical program elements so that the model's reasoning process aligns more closely with the true data and control flows in the program. By reallocating attention toward these semantically meaningful structures, the model can better reconstruct the logical relationships embedded in the program despite the absence of meaningful lexical~cues.
}


We evaluate {\tool} with multiple LLMs and code understanding datasets. Our results show that attention reallocation significantly improves LLM output prediction and execution trace prediction performance on obfuscated code, recovering much of the accuracy lost due to obfuscation, even with the accuracy comparable to that obtained on the original unobfuscated code.

Interestingly, in a few scenarios, the steered models even outperform their accuracy on the original programs. A deeper analysis reveals that the {\color{custom-blue}{\em steering process induces longer and more meaningful reasoning traces in the models}}, suggesting that they allocate more tokens to intermediate reasoning steps. This behavior resembles the {\color{custom-blue}{\em cognitive shift observed in humans}}: once lexical shortcuts such as meaningful names are removed, both humans and LLMs appear to rely more heavily on deliberate reasoning processes analogous to System 2 thinking as in humans~\cite{nguyen2026effectcodeobfuscationhuman}. Consequently, the model may analyze program semantics more thoroughly than when relying on superficial cues in the original code. 
Importantly, we showcase the practical utility of {\tool} through a case study on {\bf \em buffer overflow vulnerability analysis}, underscoring its promise for malware analysis.
This paper makes the following contributions:

\begin{enumerate}[nosep = yes, leftmargin=12pt]
    
\item Our findings have important {\bf implications} for both software security and model code intelligence. \underline{First}, they suggest that  LLMs are vulnerable to obfuscation in a way similar to humans, implying that {\em existing obfuscation techniques may remain effective against LLMs}. \underline{Second}, our results show that targeted attention steering can mitigate this limitation and substantially improve reasoning over obfuscated code. Thus, {\em steered models are useful for understanding obfuscated code in malware analysis}.

\item {\color{custom-blue}We propose a novel attention reallocation framework that relies on light-weight static priors including static slicing, data/control dependencies, and calling relations. It guides a model to focus on semantically relevant program elements for better comprehension of obfuscated code in dynamic behaviors.}


\item We demonstrate through extensive experiments across multiple LLMs and datasets that steering significantly mitigates the negative impact of obfuscation and improves model performance.

\end{enumerate}




%% file: sections/motivation-example.tex
\section{Motivating Example}
\label{sec:motivation}



{\em Non-obfuscated Code.} We start from a  Java program that computes the binomial coefficient $C(3,2)$ (Figure~\ref{fig:original}). We first modified that program by using an obfuscation tool to rename the variables, add dead code, and change the control flow (Figure~\ref{fig:full}). Both programs produce the same output, yet our goal is to investigate whether obfuscation changes \emph{how} the model gets to that answer.



We used the following to prompt {\color{custom-blue}Qwen2.5 7B} on the original code: \emph{``You will analyze the Java program provided below. First, explain how you will predict the output for the code snippet. Then, simulate it as a compiler would and print the exact console~output.''}


\begin{tcolorbox}[
  colback=black!2,
  colframe=black!15,
  boxrule=0.35pt,
  arc=1mm,
  left=1mm,right=1mm,top=0.7mm,bottom=0.7mm,
  breakable
]
\footnotesize
\textbf{Color Codes:} The colored emphasis is added by us to highlight the reasoning pattern expressed in the model generation:
\tracecue{semantic retrieval / use of pretrained algorithm knowledge};
\tracecf{control-flow inspection and simplification of obfuscation artifacts};
\tracestate{state tracking and arithmetic simulation};
\tracefinal{final output prediction}.
\end{tcolorbox}

\input{sections/example1}
\input{sections/example2}

The output from GPT-4 was shown in Figure~\ref{fig:original}. As seen, the model exploits the descriptive method name \code{binomial\-Coeffi\-cient} and answers in a retrieval-style manner, immediately~mapping the task to a familiar mathematical formula and {\em computing the output via the formula}. That is, the original version enables algorithm recognition,  encourages a \emph{semantic shortcut}. The model might fall into the scenario of not spending much space reasoning branches or narrating loop iterations; instead, it quickly maps the method name \code{binomialCoefficient} to a known concept and explains the output at the formula level. Thus, the explanation is primarily driven by recognizing the lexical cue in \code{binomialCoefficient} as a familiar combinatorial routine, with relatively little attention devoted to reasoning about statements and branch/loop conditions.

\vspace{3pt}
{\em Obfuscated Code}. Next, we used the same prompt on the~obfuscated version
with renamed variables, dead code, and changed control flow (Figure~\ref{fig:full}). As seen in its textual output,
the descriptive cue disappears and the surrounding code becomes cluttered with opaque predicates and no-op statements. Once the descriptive method name is removed and obfuscation artifacts are inserted, the model then appears to shift into an execution-style mode and the generation changes noticeably. Instead of jumping directly to a formula-level explanation as earlier, the model {\em walks through opaque predicates, dead blocks, and arithmetic updates in each iteration}. The answer is correct, but the \emph{path} to the answer is now explicitly trace-based, even though these steps are unnecessary once the algorithm is recognized. This behavior parallels the cognitive shift observed in humans: when lexical shortcuts (meaningful identifier names) are removed, LLMs rely more heavily on deliberate analytical reasoning. This phenomenon
is similar to humans' behaviors in which human developers switch to System 2's deliberate thinking when the lexical patterns in the code disappear~\cite{nguyen2026effectcodeobfuscationhuman}.

%% file: sections/example1.tex
\begin{figure}
\captionsetup{type=figure}
\examplebanner{RoyalBlue}{Example 1: Non-obfuscated program}
\begin{lstlisting}[style=motivationjava]
public class BinomialCoefficient {
    public static void main(String[] args) {
        int n = 3;
        int k = 2;
        int coefficient = binomialCoefficient(n, k);
        System.out.println(coefficient);
    }
    public static int binomialCoefficient(int n, int k) {
        int res = 1;
        if (k > n - k)
            k = n - k;
        for (int i = 0; i < k; ++i) {
            res *= (n - i);
            res /= (i + 1);
        }
        return res;
    }
}
\end{lstlisting}
\begin{tcolorbox}[
  title={Representative model generation},
  colback=RoyalBlue!2,
  colframe=RoyalBlue!35,
  boxrule=0.35pt,
  arc=1mm,
  left=1mm,right=1mm,top=0.7mm,bottom=0.7mm,
  fonttitle=\bfseries,
]
\footnotesize
\tracecue{''The \texttt{binomialCoefficient} function calculates the binomial coefficient $C(n,k)$, which is the number of ways to choose $k$ elements from a set of $n$ elements.}
\tracecue{The function uses an iterative approach to calculate the coefficient. It first checks whether $k$ is greater than $n-k$ and, if so, sets $k=n-k$ to reduce the number of multiplications.}
The result is initialized to 1, and for each $i$ the function multiplies by $(n-i)$ and divides by $(i+1)$.
\tracecue{The expected output of the program is 3 because the binomial coefficient $C(3,2)$ is 3.}
\tracefinal{Therefore, the exact console output is \texttt{3}.''}
\end{tcolorbox}
\vspace{-12pt}
\caption{Model reasoning on non-obfuscated code}
\label{fig:original}
\end{figure}


%% file: sections/example2.tex
\begin{figure}
\begin{minipage}{\linewidth}
\captionsetup{type=figure}
\examplebanner{BurntOrange}{Example 2: Fully obfuscated program}
\begin{lstlisting}[style=motivationjava]
public class C_7a2cbda3 {
    public static void main(String[] args) {
        if (((3 * 3) - 9) != 0) {
            int _obf_t5_main_405592 = 0;
            _obf_t5_main_405592++;}
        if (((2 * 2) + 1) > 0) {
            int _obf_t5_main_405592_m = 0;
            _obf_t5_main_405592_m += 0;
        } else {
            int _obf_t5_main_405592_e = 1;
            _obf_t5_main_405592_e -= 1;}
        int _obf_t5_main_405592_h = (1 ^ 1);
        if (_obf_t5_main_405592_h != 0) {
            _obf_t5_main_405592_h += 1;}
        int n = 3;
        int k = 2;
        int coefficient = m_O0lI1(n, k);
        System.out.println(coefficient);}
    public static int m_O0lI1(int n, int k) {
        int _obf_t5_m_O0lI1_591095 = 0; _obf_t5_m_O0lI1_591095 += 0;
        int res = 1;
        if (k > n - k)
            k = n - k;
        for (int i = 0; i < k; ++i) {
            res *= (n - i);
            res /= (i + 1);}
        return res;}
}
\end{lstlisting}
\begin{tcolorbox}[
  title={Representative model generation},
  colback=BurntOrange!3,
  colframe=BurntOrange!40,
  boxrule=0.35pt,
  arc=1mm,
  left=1mm,right=1mm,top=0.7mm,bottom=0.7mm,
  fonttitle=\bfseries,
]
\footnotesize
\tracecf{``The main method starts by evaluating the condition \texttt{((3 * 3) - 9) != 0}. This simplifies to $9-9 \neq 0$, i.e., $0 \neq 0$, so the first block is skipped.}
\tracecf{The next condition \texttt{((2 * 2) + 1) > 0} becomes $5>0$, so the second block executes, although it has no effect on the final value.}
\tracecf{The variable \texttt{\_obf\_t5\_main\_405592\_h} is initialized as \texttt{1 \^{} 1}, which is 0, so the following guarded block is also skipped.}

The variables $n$ and $k$ are initialized to 3 and 2, and \texttt{m\_O0lI1(n, k)} is called.
\tracestate{Inside \texttt{m\_O0lI1}, the result \texttt{res} starts at 1. Since $k > n-k$ ($2 > 1$), the method updates $k$ to $n-k=1$.}
\tracestate{The loop now runs once. For $i=0$, \texttt{res *= (n-i)} changes \texttt{res} from 1 to 3, and \texttt{res /= (i+1)} keeps it at 3.}
\tracestate{The method returns 3, which is assigned to \texttt{coefficient}.}
\tracefinal{The program prints \texttt{3}.''}
\end{tcolorbox}
\vspace{-9pt}
\caption{Model reasoning on fully obfuscated code}
\label{fig:full}
\end{minipage}
\end{figure}

%% file: sections/preliminary-study.tex
\vspace{-6pt}
\section{Preliminary Study}
\label{sec:preliminary}

We first conducted a preliminary study that establishes our empirical premise on {\em whether semantically preserving obfuscation reduces LLM performance on program output prediction even when the underlying program behavior is unchanged}. We intentionally restrict this preliminary study to the \emph{output-prediction} setting. If a model produces an incorrect output for an obfuscated program, it indicates that the model has failed to capture the correct execution.



\vspace{3pt}
{\em Data Collection and Benchmark Selection}.
Our preliminary study uses Java benchmark instances derived from two benchmarks: HumanEval~\cite{humanevalx} and CruxEval~\cite{cruxevalx}. 
{\color{custom-blue}We chose them because they provide {\em controlled, executable benchmarks} for evaluation with {\em reliably built ground truth for dynamic reasoning tasks} including output prediction and execution-trace prediction.}
For HumanEval-X~\cite{humanevalx}, we use the Java split containing 164 tasks. For CruxEval-X~\cite{cruxevalx}, we start from the benchmark's output-prediction setting and retain only those Java-translated instances whose benchmark harnesses execute reliably and from which output-prediction supervision can be derived directly. After filtering, the evaluation corpus contains 164 HumanEval-X snippets and 698 CruxEval-X snippets.


Note that the effective size of the dataset is determined not only by the number of snippets, but also by how many output-checking cases can be derived from each snippet,
i.e., how many usable seed checks each snippet provides and how many variants of checks can~be generated from those seeds. 
HumanEval-X contains fewer snippets overall, but many of its benchmark harnesses include multiple boolean checks.
Each such check can serve as a seed assertion and can further produce additional validated counterfactual mutations, resulting in relatively large case packs per snippet. 

\vspace{3pt}
{\em Counterfactual Output-Prediction Case Construction}. We derive output-prediction labels automatically from the original benchmark harnesses rather than annotating them manually. For each retained snippet, we construct a validated counterfactual case pack starting from seed true-cases extracted from the harness. In HumanEval-X, these seeds correspond to boolean expressions inside \code{List<Boolean> correct = Arrays.asList(...)}; in CruxEval-X, they correspond to \code{assert(...)} predicates. Each seed is {\em re-executed} on the actual program, and seeds that do not evaluate to \code{true} are discarded as unstable. Starting from the surviving seeds, we generate candidate false-cases by applying small mutations to the original checks.
{\color{custom-blue}Following mutation testing, our mutation operation types include \code{expected\_value\_mutation}, \code{threshold\_mutation}, \code{literal\_per\-turbation}, \code{argu\-ment\_perturbation}, \code{comparator\_flip}, and \code{nega\-tion\_fallback}.}
Each mutated case is retained only if it executes successfully and evaluates to \code{false}. {\color{custom-blue} Finally, we remove trivial literal cases, including 
\code{assert true/false}, \code{assert stringLiteral}, as the model can easily guess the output after mutations. We keep only non-trivial cases that are validated through program execution.}
This procedure yields 1,922 validated output-prediction cases for HumanEval-X and 1,378 for CruxEval-X. Averaged over snippets, this corresponds to 11.72 cases per HumanEval snippet and 1.97 cases per CruxEval-X snippet. Although CruxEval-X contributes more programs, HumanEval-X contributes substantially larger case packs per program.

\vspace{2pt}
{\em Obfuscation Augmentation and Semantic Validation.}~We augment each clean Java snippet with semantics-preserving source-to-source obfuscated variants. Our transformation space is organized into four representative families: {\em identifier renaming}, {\em dead-code injection}, {\em control-flow flattening}\cite{control-flow-flattening}, and {\em call indirection}. To ensure that obfuscation does not change program semantics, each transformed variant is validated against the original benchmark harness and retained only if the observable behavior matches the original, including task-relevant labels under the harness.

\vspace{2pt}
\emph{Experimental Procedure and Metric.}
For each individual case in a case pack, we prompt the model separately with the corresponding program and ask it to return the answer containing a binary prediction, \texttt{T} or \texttt{F}. Each case is evaluated with {\em three independent runs} under the same prompting and decoding regime. 
A case is counted as \emph{passed} at $k$ if at least one of the first $k$ independent responses for that case matches the ground-truth label. The reported score is the proportion of cases that pass within $k$ attempts. 



\begin{table}[t]
\centering
\footnotesize
\setlength{\tabcolsep}{3pt}
\caption{Preliminary output-prediction results. Obfuscated results aggregate across all 4 obfuscation types.}
\label{tab:preliminary}
\vspace{-9pt}
\begin{tabular}{llcc}
\toprule
& & P@1 (Orig.) (\%) & P@1 (Obf.) (\%)\\
\midrule
\multirow{2}{*}{Qwen2.5-7B}
 & HumanEval-X & 76.49 & 64.01 \\
 & CruxEval-X  & 83.11 & 71.40 \\
\midrule
\multirow{2}{*}{DeepSeek-6.7B}
 & HumanEval-X & 77.74 & 68.29 \\
 & CruxEval-X  & 78.98 & 66.93 \\
\midrule
\multirow{2}{*}{Qwen2.5-14B}
 & HumanEval-X & 94.63 & 85.31 \\
 & CruxEval-X  & 92.14 & 86.48 \\
\midrule
\multirow{2}{*}{DeepSeek-V2-Lite}
 & HumanEval-X & 90.31 & 80.18 \\
 & CruxEval-X  & 91.56 & 81.11 \\
\bottomrule
\end{tabular}

\end{table}


\vspace{1pt}
\emph{Preliminary Results.}
In Table~\ref{tab:preliminary}, the obfuscated condition is consistently lower than the original condition at Pass@1. 
This pattern appears on both HumanEval-X and CruxEval-X, indicating that semantics-preserving obfuscation degrades output-prediction performance. This result has the following implications: 

(1) LLM performance {\bf degrades substantially} when code obfuscation is present. In particular, their ability to reason about program behavior becomes significantly weaker on obfuscated programs, mirroring the confusion experienced by humans when reading obfuscated code as reported by Nguyen {\em et al.}~\cite{nguyen2026effectcodeobfuscationhuman}.

(2) The decline in accuracy arises because obfuscation disrupts surface-level lexical and structural cues that LLMs often exploit when analyzing non-obfuscated code, (see motivating example).


(3) These findings motivate our next question in this work:~whether inference-time attention steering can redirect an LLM toward semantically relevant program elements and thereby (partially) recover the performance lost under obfuscation. This would help improve its code understanding, leading to better malware analysis.


%% file: sections/technical.tex
\section{Post-Hoc Attention Steering}
\label{sec:method}

In this work, {\color{custom-blue}we evaluate models' capability in two fundamental tasks in code understanding: \emph{execution prediction} and \emph{output prediction}. Execution prediction requires reasoning about which statements are executed for an input, while output prediction requires the computation of variables' values, deciding the final result. 

To guide the model without training/fine-tuning, {\tool} constructs a static relevance prior over code tokens using lightweight program analysis. These signals are not intended to pre-solve these above tasks that require dynamic reasoning; rather,  the key intuition is that these static signals provide a soft bias that helps the model allocate more attention to code regions that are structurally connected to the program behavior being reasoned about when the lexical cues disappear in the source code.


Concretely, {\tool} uses control-flow and dependence information to estimate
semantic relevance. Control-flow information captures how statements may be
ordered and guarded during execu\-tion, while dependence information captures how
values may flow through assignments, variables, calls, predicates, and return
expressions. 
%
Based on this analysis, {\tool} maps relevant program elements to their
corresponding source spans and then to prompt-token positions. The resulting
token set forms a soft attention prior used by the steering algorithm during
decoding, while the full program remains available in the context.
Importantly, this prior does not mask irrelevant code or remove
obfuscation artifacts.} 
In fact, this prior is static, it may be incomplete or noisy for concrete runtime tasks; the LLM must still reason about branch outcomes, loop iterations, intermediate values, and final outputs during generation. 




\subsection{Program Slicing to Build Steering Priors}

While the statements along the {\color{custom-blue}static control flow} from the input are straightforward to identify, let us explain the procedure that we use {\color{custom-blue}static backward slicing} to identify
the crucial tokens for output prediction. First, the {\bf steering prior} is a prompt-aligned relevance distribution derived from our program dependence approximation. It provides a stable token-level signal indicating which parts of the program are likely to affect the output prediction.

\vspace{1pt}
{\em Data Pre-processing.} 
For a program, we first identify a set of observable \emph{task sinks} $S$, defined as the expressions whose~evaluated values determine the task outcome. 
As a correctly predicted output is crucial for dynamic reasoning,
$S$ consists of the expressions passed to \code{System\-.out\-.print\-ln\-(...)} within the \code{main} method. In harness-based benchmarks, $S$ consists of the boolean predicates that encode correctness, such as assertions or equality checks (e.g., \code{s.vowels\-Count\-("ACEDY") == 3}). For each sink $s \in S$, we identify the program elements that directly contribute to its value. If the value of $s$ depends on a call to a user-defined method, we include the corresponding call result and the associated return expressions as \emph{output seeds}, and designate that callee as a \emph{target method}. Otherwise, the enclosing method of $s$ is treated as the target method. When multiple sinks or multiple target methods are present, we retain all of them and aggregate their reachable dependencies, rather than imposing a single sink. This ensures that the subsequent analysis captures all elements that may affect the observable outcome.

{\em Building Static Program Slices.} Starting with a target method and any helper methods that can be reached from them, we build a {\color{custom-blue}static dependence graph} using Joern. We only keep the parts of the graph that are necessary for steering, and represent it as
\begin{equation}
G=(V, E_{\mathrm{data}} \cup E_{\mathrm{ctrl}} \cup E_{\mathrm{call}}),
\label{eq:graph}
\end{equation}
where the nodes correspond to program elements such as assignments, variable uses, control conditions, return expressions, function parameters, method calls, and output-related predicates. The edges capture different types of relationships: data edges represent how values flow between variables (i.e., definition--use relations), control edges represent how conditions (e.g., if-statements) determine if other statements execute, and call edges connect arguments to parameters and link return values back to where they are used. 


\vspace{2pt}
{\em Identifying crucial tokens for a steering prior}. Our goal is to identify the subset of program elements relevant to predicting the output for a given input. Considering only backward dependencies from the output may include statements that do not depend on the input, while considering only forward dependencies from the input may include computations that do not contribute to the output. To isolate the relevant computation, we focus on the intersection of these two views---i.e., {\em the program elements that both depend on the input and influence the output}. This {\em intersection} approximates the core input-to-output dependency path of the program and provides a principled basis for guiding attention toward more relevant tokens.

Let $I \subseteq V$ denote the \emph{input seeds}, consisting of the formal parameters of each target method together with the actual arguments at all call sites that can reach a task sink. Let $O \subseteq V$ denote the \emph{output seeds}, including the task sinks in $S$ and any return or call-result nodes whose values flow into those sinks. In a simple \texttt{main}-style program, $O$ includes the printed expression and any callee return that contributes to it. In a harness-based setting, $O$ includes each assertion or equality predicate and any user-defined call results appearing within those predicates. Thus, $O$ represents the set of observable values that determine the final answer.

To identify relevant program elements, we first compute the backward-reachable set from the output seeds,
\begin{equation}
R^{-}(O)=\{v\in V \mid \exists o\in O:\ v \rightsquigarrow o\},
\end{equation}
which captures nodes that may influence the output. However, this set may include statements that are independent of the input. We also compute the forward-reachable set from the input seeds,
\begin{equation}
R^{+}(I)=\{v\in V \mid \exists i\in I:\ i \rightsquigarrow v\},
\end{equation}
which captures nodes that depend on the input but may not affect the output.
We then take the intersection of these two sets,
\begin{equation}
V_{\mathrm{slice}} = R^{+}(I) \cap R^{-}(O),
\end{equation}
to retain only the nodes that lie on dependency paths from inputs to outputs. This step filters out unrelated computations and yields a compact set of program elements that are both input-dependent and output-relevant. Finally, we include the governing control predicates of nodes in $V_{\mathrm{slice}}$ to ensure that branch conditions affecting the execution of output-relevant statements are preserved. We refer to the resulting set as a {\bf \em slice prior}: a slice-inspired relevance approximation designed to guide attention. This prior enables our steering mechanism to bias the model toward output prediction.


\subsection{Token Weighting and Normalization}
\label{sec:token-weight}

Our method adapts PASTA~\cite{PASTA}, an existing training-free, head-local attention-steering procedure to our problem. We replace user-highlighted text spans with a statically constructed relevance prior over code tokens, derived by {\color{custom-blue} static} control flow analysis or program slicing, and we apply the steering update only to a sparse, automatically calibrated subset of heads during decoding~\cite{Towards-Automated-Circuit-Discovery-for-Mechanistic-Interpretability}.
Let $x_{1:n}$ denote the prompt tokens, consisting of the task instruction followed by a  program, and let $y_{1:T}$ denote the generated answer. At decode step $t$, decoder layer $\ell$, and attention head $h$, let
\begin{equation}
A^{(t)}_{\ell,h}(k)=\operatorname{softmax}\!\left(S^{(t)}_{\ell,h}\right)_k,
\label{eq:base_attn}
\end{equation}
be the attention probability assigned to valid key position $k\in\{1,\dots,K_t\}$. Our goal is to bias part of this attention mass toward prompt positions that are more likely to 
be relevant to model reasoning on dynamic behaviors,
while leaving model weights, tokenization, and the autoregressive decoding algorithm unchanged.

Each node $v\in V_{\mathrm{slice}}$ (computed earlier) is mapped to its span in the source code and then to the prompt tokenization used by the model. Let $\Pi(v)$ denote the set of prompt-token indices aligned to node $v$. Before decode-time mixing, we assign zero mass to instruction tokens outside the code region and concentrate the prior on prompt tokens aligned to selected code nodes. This projection is tokenizer-dependent; when identifiers split into subword pieces, the induced prior may blur statement boundaries.

Each selected node receives a nonnegative weight $\omega(v)$ based on a set of conservative heuristics, including node type, proximity to the output seed, and whether the node lies on multiple input-to-output paths. These weights are accumulated on prompt tokens: 
\begin{equation}
 w_k = \sum_{v\in V_{\mathrm{slice}}: k\in\Pi(v)} \omega(v) + \lambda\,\mathbf{1}[k\text{ belongs to a target-method body}],
\label{eq:token_weight}
\end{equation}
where the optional baseline term $\lambda\ge 0$ prevents the prior from collapsing onto only a few tokens. The {\bf static prompt prior} is
\begin{equation}
\bar p(k)=\frac{w_k}{\sum_{j=1}^{n} w_j}.
\label{eq:static_prior}
\end{equation}
This prior is designed to encode likely semantic relevance,
not capturing a complete dependence analysis for all program elements.

\subsection{Decode-Time Prior and Steering Update}
The static prior in Equation~\ref{eq:static_prior} is defined over prompt tokens, but generation unfolds over time and later decode steps also attend to previously generated tokens. To let the intervention evolve over generation without changing its basic form, we partition the continuation into equal-count temporal bins. The purpose of these bins is to let the prior emphasize different parts of the context at different phases of generation, for example, stronger prompt anchoring early in the explanation and more tolerance for generated context later.

Let $b(t)$ be the bin containing decode step $t$, let $\bar p^{(b)}$ denote the prompt prior used in bin $b$ (in the simplest case, $\bar p^{(b)}=\bar p$ for all $b$), and let $r^{(t)}$ denote a recency prior over valid keys at step $t$. We define the decode-time prior as
\begin{equation}
p^{(t)}=
\begin{cases}
\bar p^{(b(t))}, & t=1,\\
\operatorname{Normalize}\!\left((1-\rho_{b(t)})\bar p^{(b(t))}+\rho_{b(t)} r^{(t)}\right), & t>1,
\end{cases}
\label{eq:decode_prior}
\end{equation}
where $\rho_{b(t)}\in[0,1]$ controls how strongly the current bin mixes prompt relevance with recency.
For each steered layer $\ell$ and head $h$, we then reweight the already normalized attention distribution using this decode-time prior:
\begin{equation}`
\widetilde A^{(t)}_{\ell,h}(k)=
\frac{A^{(t)}_{\ell,h}(k)\left(p^{(t)}_k+\varepsilon\right)^{\beta}}
{\sum_{j=1}^{K_t} A^{(t)}_{\ell,h}(j)\left(p^{(t)}_j+\varepsilon\right)^{\beta}},
\label{eq:steer}
\end{equation}
where $\beta>0$ controls steering strength and $\varepsilon>0$ is a small numerical-stability constant. Equation~\ref{eq:steer} redistributes existing attention mass toward positions with higher slice prior; it does not add external tokens or update model parameters. 

\subsection{Sparse Head Calibration}
{\em We do not steer every attention head. Instead, steering is restricted to a sparse subset of heads in each steered layer}~\cite{in-context-learning-and-induction-heads, Interpretability-in-the-Wild:-A-Circuit-for-Indirect-Object-Identification-in-GPT-2-Small}. This is both practical and methodological: the intervention should be strong enough to shift attention meaningfully, but narrow enough to avoid broad collateral disruption of unrelated attention patterns.
We choose this subset automatically using a small calibration set $\mathcal{C}$. For each layer-head pair $(\ell,h)$, we collect the attention distribution at the \emph{first decode step} and score its alignment with the prompt prior: 
\begin{equation}
g_{\ell,h}=\mathbb{E}_{x\sim\mathcal{C}}\left[\sum_{k=1}^{n_x} A^{(1)}_{\ell,h}(k)\,\bar p_x(k)\right].
\label{eq:head_score}
\end{equation}
We use the first decode step as it provides a stable prompt-conditioned attention profile before  generated tokens begin to dominate the context. Heads with larger $g_{\ell,h}$ already attend more to slice-relevant prompt positions and are the most promising targets for steering.

Let $M_{\ell,h}\in\{0,1\}$ denote the binary mask obtained by retaining only the highest-scoring fraction of heads in each steered layer. In the reported experiments, this fraction is held fixed within each model configuration and is roughly one quarter of the heads in each steered layer in our main setup; exact numerical values are deferred to Section~4. The final attention distribution is
\begin{equation}
\widehat A^{(t)}_{\ell,h} = A^{(t)}_{\ell,h} + M_{\ell,h}\left(\widetilde A^{(t)}_{\ell,h} - A^{(t)}_{\ell,h}\right).
\label{eq:masked_update}
\end{equation}
If $M_{\ell,h}=0$ for a head, or if $\beta=0$, the steering operator reduces to the identity transformation for that head.

\subsection{Decode-Time Steering Intervention}
The intervention is applied only during autoregressive decoding, not during prompt prefill. This separation is important because our goal is to influence how the model \emph{uses} the prompt while generating the explanation, rather than to alter the initial prompt encoding itself. In practice, we therefore build the KV cache from a prompt prefix without steering and begin generation from the final prompt token, so that the first steered step corresponds to a single-token query attending over the full prompt context.

{\color{custom-blue}
To localize the intervention, we apply steering only in a late band of
decoder layers, close to token selection. This provides enough influence
near generation while keeping earlier representation formation largely untouched,
thereby reducing collateral disruption to unrelated attention patterns and making
the effect easier to attribute. Across all reported runs, the layer band,
steering schedule, and steering strength are held fixed, while the active head
subset is selected automatically by the calibration~criterion in
Equation~\ref{eq:head_score}}. 





%% file: sections/emperical.tex
\section{Empirical Evaluation}
\label{sec:eval}



We seek to answer the following research questions:

\textbf{RQ1. Effectiveness of Model Steering.}
To what extent does our steering method recover LLM performance on obfuscated code?

\textbf{RQ2. Sensitivity to Obfuscation Type.}
Which obfuscation techniques induce the most degradation in model performance, and which settings are most recoverable by steering?

\textbf{RQ3. Behavioral Mechanisms of Obfuscation and Steering.}
How do obfuscation and steering alter both the reasoning trace as well as the model's internal attention dynamics during generation?
\\
\hspace*{1.5em}\textbf{RQ3.1 Reasoning Trace-Level Behavior.}
How do obfuscation and steering change the quality and form of reasoning traces?
\\
\hspace*{1.5em}\textbf{RQ3.2 Attention-Level Behavior.}
With obfuscated code, what head-level attention patterns distinguish reasoning-first from direct-answer outputs, and how are these patterns reshaped by steering?

{\color{custom-blue}
\textbf{RQ4. Generalization to Other Obfuscation Types.}
How well can {\tool} be generalized to other obfuscation types?
}

%% file: sections/RQ1-new.tex
\section{Effectiveness of Model Steering (RQ1)}
\label{sec:rq1}

\input{tables/output-prediction-table-2}



\subsection{Output Prediction}
\label{sec:rq1-output}

\subsubsection{Dataset, Procedure, and Metrics}
\label{sec:rq1-output-setup}

For the output-prediction task, we reuse the same benchmark construction, prompting protocol, and case-level evaluation setup as in Section~\ref{sec:preliminary}. 
%
{\color{custom-blue}
We also used the same datasets as in Section~\ref{sec:preliminary}: HumanEval~\cite{humanevalx} and CruxEval~\cite{cruxevalx}. 
The programs in those two datasets allow us to isolate the effect of attention steering without confounding factors such as build failures, malware setup, missing dependencies, or unreliable harnesses.
We chose our models as in Table~\ref{tab:main-eval} as they are strong open-weight code models with support for attention extraction and inference-time steering. Moreover, they were released before the chosen datasets, reducing the data-contamination risk. 
}
Each case is prompted and evaluated with three independent runs under the same decoding regime.
The primary output-prediction metric remains \textbf{Pass@}$k$.
To assess the effectiveness of {\tool}, we define \textbf{Restoration Ratio} in attempt budget $k$ as
\begin{equation}
R_P@k
=
\frac{P_{\text{Obfuscated+Steering}}@k - P_{\text{Obfuscated}}@k}
     {P_{\text{Original}}@k - P_{\text{Obfuscated}}@k}
\times 100 
\label{eq:restoration-output}
\end{equation}
{\em This measures the fraction of the obfuscation-induced performance drop that is recovered by steering}. 
$100\%$ means full recovery to the original code, values above $100\%$ mean \emph{over-recovery} (i.e., the steered obfuscated condition exceeds the original one), and negative values mean that steering hurts performance relative to the obfuscated~one.

\subsubsection{Empirical Result}

As seen in Table~\ref{tab:main-eval}, under steering, \code{Qwen2.5-7B} reaches 77.38--90.32\% on HumanEval-X and 80.43\%/ 89.14\% / 91.83\% on CruxEval-X; \code{DeepSeek-6.7B} reaches 74.89--89.11\% and 70.25--89.63\%.
That is, even under obfuscation, {\tool} allows all four models to maintain strong output-prediction performance, with the strongest absolute results obtained by \code{Qwen2.5-14B} and the weakest by \code{DeepSeek-6.7B}.
As seen in Pass@1, steering improves first-attempt obfuscated performance in every setting.
These Pass@1 numbers gain translates into a substantial recovery of the obfuscation-induced loss. At Pass@1, {\tool} surpasses with 107.1\% of the performance for \code{Qwen2.5-7B} on the original code in HumanEval-X, recovers 69.8\% of the lost performance for \code{DeepSeek-6.7B}, 88.3\% for \code{Qwen2.5-14B}, and 70.7\% for \code{DeepSeek-V2-Lite}. On CruxEval-X, the restoration ratios are 77.1\%, 27.6\%, 81.1\%, and 48.6\%. That is, it not only improves obfuscated performance in absolute terms, but in many cases recovers a large fraction of the original-code gap, and in one setting (\code{Qwen2.5-7B}, HumanEval-X) even exceeds the original-code baseline.

Comparing across models, two complementary patterns emerge. First, stronger models remain stronger after steering: \code{Qwen2.5-14B} has the best absolute performance in every setting, followed by \code{DeepSeek-V2-Lite}, while \code{Qwen2.5-7B} and \code{DeepSeek-6.7B} remain lower in absolute accuracy. Second, the largest \emph{relative} recovery does not always occur in the strongest model. 
In contrast, stronger models already preserve more useful semantics under obfuscation, so a fixed steering configuration has less remaining error to correct, yielding smaller relative gains even when accuracy remains highest. Pass@2 and Pass@3 follow the same qualitative trend as Pass@1.

{\color{custom-blue} 
We also compared {\tool} against the prompting approach~in which we prompt a model under study to focus its attention on the set of collected code tokens as described in Section~\ref{sec:token-weight}. As~seen in Table~\ref{tab:main-eval}, naive prompting is ineffective or less effective than {\tool}. Pass@1 restoration ratios with prompting are lower in 7/8 rows or approximately the same in one row as {\tool}.


To show its potential scalability, we chose the programs in DeepMind Code Contests dataset~\cite{li2022competition} with 300-600 LOCs (10$\times$-20$\times$ longer than an average program in HumanEval-X and CRUXEval-X). Among all 820 programs, Joern was able to
successfully perform slicing on 152 of them. Running Qwen2.5-7B on those programs, obfuscation reduced  Pass@1 from 36.0\% to 33.9\%, and {\tool} recovered it to 35.8\%, yielding a 87.8\% restoration~rate. Model steering incurs no token cost. The maximum program size that {\tool} can process is bounded by the underlying model's context length.

}

\begin{table}[t]
\centering
\small
\caption{{\tool}'s effectiveness on execution prediction}
\label{tab:trace-prediction}
\vspace{-9pt}
\tabcolsep 1.2pt
\begin{tabular}{l|l|cc|cc|cc}
\toprule
\multirow{2}{*}{Model} & \multirow{2}{*}{Dataset} & Orig. & \shortstack{Obf. no \\ Pr or S} & \shortstack{Obf. w/\\ Prompt} & \shortstack{Restor.\\Ratio} &
\shortstack{Obf. w/ \\ Steer} & \shortstack{Restor.\\Ratio}\\
\cmidrule(lr){3-3}\cmidrule(lr){4-4}\cmidrule(lr){5-5}\cmidrule(lr){6-6}\cmidrule(lr){7-7}\cmidrule(lr){8-8}
& & F1 & F1 & F1 & $R_{Pr} (\%)$ & F1 & $R_{St} (\%)$ \\
\midrule
\multirow{2}{*}{Qwen2.5-7B}
 & HumanEval & 36.52 & 33.12 & {\color{custom-blue}27.82} & {\color{custom-blue}-155.9} & 34.78 & 48.82 \\
 & CruxEval  & 43.35 & 39.32 & {\color{custom-blue}38.49} & {\color{custom-blue}-20.6} & 42.29 & 73.70 \\
\midrule
\multirow{2}{*}{Qwen2.5-14B}
 & HumanEval & 40.14 & 38.17 & {\color{custom-blue}36.31} & {\color{custom-blue}-94.6} & 39.74 & 79.70 \\
 & CruxEval  & 49.20 & 45.88 & {\color{custom-blue}45.6} & {\color{custom-blue}-8.66} & 47.94 & 62.05 \\
\bottomrule
\end{tabular}
\end{table}

\subsection{Execution-Trace Recovery}
\label{sec:rq1-trace}
\subsubsection{Dataset, Procedure, and Metrics}
\label{sec:rq1-trace-setup}


We use the same dataset as presented in Section~\ref{sec:preliminary}, which contains
3,300 output-prediction cases in total. After removing 58 cases in HumanEval due to non-terminating execution, we obtain 3,242 validated execution-trace cases in total.
We executed each program in the three settings (\emph{original}, \emph{obfuscated}, and \emph{obfuscated code with Steering}) with its input~and collected the execution trace. We compute \textbf{F1} against the oracle using longest-common-subsequence overlap. Let $L=\mathrm{LCS}(\hat{y}, y)$ be the longest common subsequence between the predicted trace $\hat{y}$ and the oracle trace $y$. We define
$P_{\mathrm{trace}}=\frac{L}{|\hat{y}|},
R_{\mathrm{trace}}=\frac{L}{|y|},
F1_{\mathrm{trace}}=\frac{2P_{\mathrm{trace}}R_{\mathrm{trace}}}{P_{\mathrm{trace}}+R_{\mathrm{trace}}}.$
This gives partial credit when a model recovers the correct execution order partially. 
The F1 restoration ratio is defined~as
\begin{equation}
R_{\text{F1}}
=
\frac{\text{F1}_{\text{trace(Obfuscated+Steering)}} - \text{F1}_{\text{trace(Obfuscated)}}}
     {\text{F1}_{\text{trace(Original)}} - \text{F1}_{\text{trace(Obfuscated)}}}
\times 100
\label{eq:restoration-trace}
\end{equation}

\subsubsection{Empirical Result}
\label{sec:rq1-trace-results}

Table~\ref{tab:trace-prediction} shows that steering improves the model performance on obfuscated code in every row of Table~\ref{tab:trace-prediction}. For \code{Qwen2.5-14B}, F1 rises from 38.17 to 39.74 on HumanEval and from 45.88 to 47.94 on CruxEval, with restoration ratios of 79.70\% and 62.05\%. For \code{Qwen2.5-7B}, F1 rises from 33.12 to 34.78 on HumanEval and from 39.32 to 42.29 on CruxEval, corresponding to F1 restoration ratios of 48.82\% and 73.70\%. These results show that steering recovers part of the performance lost caused by obfuscation. 
Obfuscation also tends to add code constructs that expand the execution trace, adding more burden on a model to reason on obfuscated code.

In Table~\ref{tab:main-eval}, Qwen models provide the strongest and most stable absolute performance in output prediction, with \code{Qwen2.5-14B} achieving the best results overall and \code{Qwen2.5-7B} remaining competitive under obfuscation. 
For Qwen, the larger model remains stronger in absolute F1, while the smaller model can still show substantial recovery under steering. 
Steering recovers part of the execution semantics lost under obfuscation, but the remaining difficulty is concentrated in reconstructing the entire ordered trace and values.
{\color{custom-blue} Table~\ref{tab:trace-prediction} also shows that F1 scores with prompting on obfuscated code are lower or approximately the same as the no-prompt variant, thus largely ineffective as seen in the negative restoration ratios}.

%% file: tables/output-prediction-table-2.tex
\begin{table*}[t]
\centering
\small
\setlength{\tabcolsep}{2.5pt}
\caption{Effectiveness of {\tool} in output prediction (RQ1). Pass@k denotes case-coverage union across the first $k$ runs for the same prompt and case tuple. Obfuscated conditions aggregate over the 4-obfuscation types. Values are percentages.}
\label{tab:main-eval}
\vspace{-9pt}
\begin{tabular}{llccccccccccccccccccc}
\toprule
\multirow{2}{*}{Model} & \multirow{2}{*}{Dataset} & \multicolumn{3}{c}{\shortstack{Original\\No-Obfuscation}} & \multicolumn{3}{c}{\shortstack{Obfuscated \\ w/o Prompt/Steer}} & \multicolumn{3}{c}{\shortstack{Obfuscated\\ w/. Prompt}} & \multicolumn{3}{c}{\shortstack{Restoration\\Ratio w/ Prompt} (\%)}
& \multicolumn{3}{c}{\shortstack{Obfuscated\\ w/ Steering}} & \multicolumn{3}{c}{\shortstack{Restoration\\Ratio w/ Steer} (\%)} \\
\cmidrule(lr){3-5}\cmidrule(lr){6-8}\cmidrule(lr){9-11}\cmidrule(lr){12-14}\cmidrule(lr){15-17}\cmidrule(lr){18-20}
& & P@1 & P@2 & P@3 & P@1 & P@2 & P@3 & P@1 & P@2 & P@3 & R@1 & R@2 & R@3 & P@1 & P@2 & P@3 & R@1 & R@2 & R@3 \\
\midrule
\multirow{2}{*}{Qwen2.5-7B}

          & HumanEval & 76.49 & 86.15 & 88.82 & 64.01 & 76.93 & 82.83 & {\color{custom-blue}72.99} & {\color{custom-blue}81.74} & {\color{custom-blue}85.74} & {\color{custom-blue}72.0} & {\color{custom-blue}52.2} & {\color{custom-blue}48.5} & \textcolor{dkgreen}{77.38} & \textcolor{dkgreen}{86.90} & \textcolor{dkgreen}{90.32} & \textcolor{dkgreen}{107.1} & \textcolor{dkgreen}{108.1} & \textcolor{dkgreen}{125.0} \\
 & CruxEval & 83.11 & 92.90 & 95.34      & 71.40 & 83.16 & 87.81     & {\color{custom-blue}79.91} & {\color{custom-red}89.68}  & {\color{custom-red}95.37}     & {\color{custom-blue}72.7} & {\color{custom-red}66.9} & {\color{custom-red}100.4} &
 80.43 & 89.14 & 91.83     & 77.1 & 61.4 & 53.4 \\
\midrule
\multirow{2}{*}{DeepSeek-6.7B}
          & HumanEval & 77.74 & 84.84 & 88.40 & 68.29 & 81.43 & 86.47 & {\color{custom-blue}65.91} & {\color{custom-blue}84.52} & {\color{custom-red}90.08} & {\color{custom-blue}-25.1} & {\color{custom-blue}90.6} & {\color{custom-red}186.9} & \textcolor{dkgreen}{74.89} & \textcolor{dkgreen}{85.39} & 89.11 & \textcolor{dkgreen}{69.8} & \textcolor{dkgreen}{116.1} & 136.8 \\

 & CruxEval  & 78.98 & 88.96 & 91.98     & 66.93 & 81.63 & 88.44     & {\color{dkgreen}70.36} & {\color{custom-blue}82.75} & {\color{custom-red}91.83}     & {\color{dkgreen}28.5}  & {\color{custom-blue}15.3}  & {\color{custom-red}95.8}  & 
 {\color{dkgreen}70.25} & 84.43 & 89.63     & {\color{dkgreen}27.6} & 38.2 & 33.6 \\
\midrule
\multirow{2}{*}{Qwen2.5-14B}
          & HumanEval & 94.63 & 96.81 & 97.37 & 85.31 & 93.13 & 95.82 & {\color{custom-blue}86.82} & {\color{custom-blue}93.66} & {\color{custom-blue}96.29} & {\color{custom-blue}16.2} & {\color{custom-blue}14.4} & {\color{custom-blue}30.2} & \textcolor{dkgreen}{93.54} & \textcolor{dkgreen}{96.39} & \textcolor{dkgreen}{97.10} & \textcolor{dkgreen}{88.3} & \textcolor{dkgreen}{88.6} & \textcolor{dkgreen}{82.6} \\
          & CruxEval & 92.14 & 97.28 & 98.33 & 86.48 & 94.56 & 96.95 & {\color{custom-blue}89.42} & {\color{custom-blue}96.25} & {\color{custom-blue}98.15} & {\color{custom-blue}51.9} & {\color{custom-blue}62.1} & {\color{custom-blue}87.1} & \textcolor{dkgreen}{91.07} & \textcolor{dkgreen}{96.70} & \textcolor{dkgreen}{98.17} & \textcolor{dkgreen}{81.1} & \textcolor{dkgreen}{78.7} & \textcolor{dkgreen}{88.4} \\

\midrule
\multirow{2}{*}{DeepSeek-V2-Lite}
          & HumanEval & 90.31 & 94.78 & 96.08 & 80.18 & 90.62 & 92.37 & {\color{custom-blue}68.00} & {\color{custom-blue}82.35} & {\color{custom-blue}88.19} & {\color{custom-blue}-120.2} & {\color{custom-blue}-198.8} & {\color{custom-blue}-112.7} & \textcolor{dkgreen}{87.34} & \textcolor{dkgreen}{92.43} & \textcolor{dkgreen}{93.71} & \textcolor{dkgreen}{70.7} & \textcolor{dkgreen}{43.5} & \textcolor{dkgreen}{36.1} \\
          & CruxEval & 91.56 & 95.87 & 97.12 & 81.11 & 91.39 & 94.66 & {\color{custom-blue}79.66} & {\color{custom-blue}93.79} & {\color{custom-red}97.42} & {\color{custom-blue}-13.9} & {\color{custom-blue}53.6} & {\color{custom-red}112.1} & \textcolor{dkgreen}{86.19} & \textcolor{dkgreen}{93.92} & 95.78 & \textcolor{dkgreen}{48.6} & \textcolor{dkgreen}{56.5} & 45.5 \\

\bottomrule
\end{tabular}
\end{table*}

%% file: sections/Stratified_Result_by_Obf_type.tex
\section{Sensitivity to Obfuscation Techniques (RQ2)}
\label{sec:rq2}

In this experiment, we study which obfuscation type degrades model performance the most and which is most recoverable by our steering. We reuses exactly the same output-prediction dataset, prompting procedure, and evaluation metrics as in Section ~\ref{sec:rq1-output}. Table~\ref{tab:rq3-obf-type} shows that damage and recoverability vary substantially across obfuscation types. On HumanEval-X, the largest Pass@1 drop comes from \textit{Identifier Renaming} (76.49$\rightarrow$40.20), while \textit{Control-Flow Flattening} is the least harmful (76.49$\rightarrow$73.28). On CruxEval-X, the largest drop comes from \textit{Dead-Code Injection} (83.11$\rightarrow$66.08), whereas \textit{Call Indirection} is the least harmful (83.11$\rightarrow$77.16). Steering is most effective on \textit{Control-Flow Flattening} for HumanEval-X (73.28$\rightarrow$78.71, 169.10\%) and on \textit{Identifier Renaming} for CruxEval-X (74.51$\rightarrow$85.48, 127.50\%). This confirms our intuition of
{\em slice-based steering}. The weakest recoveries are \textit{Call Indirection} on HumanEval-X (83.06\%) and \textit{Control-Flow Flattening} on CruxEval-X (53.99\%).

This pattern matches our interpretation of obfuscation. Identifier renaming mainly disrupts lexical cues, so steering can recover performance by re-centering the model on relevant tokens rather than surface names. Dead-code injection and call indirection are less damaging to begin with, leaving less headroom for recovery.

\input{tables/stratified}

%% file: tables/stratified.tex
\begin{table}[t]
\centering
\scriptsize
\setlength{\tabcolsep}{3.5pt}
\caption{Stratified evaluation of Qwen2.5-7B model in output prediction ($pass@1$) over the four obfuscation types.}
\label{tab:rq3-obf-type}
\vspace{-9pt}
\resizebox{\columnwidth}{!}{%
\begin{tabular}{llcccc}
\toprule
\multirow{2}{*}{Dataset} & \multirow{2}{*}{Obfuscation Technique} & \multicolumn{1}{c}{Orig.} & \multicolumn{1}{c}{Obf. w/o} & \multicolumn{1}{c}{Obf. w/} & \multicolumn{1}{c}{Restor.} \\
& & & Steering & Steering & Ratio \\
\midrule
\multirow{4}{*}{H/Eval} & Dead-Code Injection & \multirow{4}{*}{76.49} & 70.54 & \textcolor{dkgreen}{76.79} & \textcolor{dkgreen}{105.00} \\
 & Call Indirection &  & 72.00 & 75.73 & 83.06 \\
 & Control-Flow Flattening &  & 73.28 & \textcolor{dkgreen}{78.71} & \textcolor{dkgreen}{169.10} \\
 & Identifier Renaming &  & 40.20 & \textcolor{dkgreen}{78.30} & \textcolor{dkgreen}{104.99} \\
\cmidrule(lr){1-6}
\multirow{4}{*}{C/Eval} & Dead-Code Injection & \multirow{4}{*}{83.11} & 66.08 & 77.15 & 64.99 \\
 & Call Indirection &  & 77.16 & 83.01 & 98.28 \\
 & Control-Flow Flattening &  & 67.84 & 76.08 & 53.99 \\
 & Identifier Renaming &  & 74.51 & \textcolor{dkgreen}{85.48} & \textcolor{dkgreen}{127.50} \\
\bottomrule
\end{tabular}
}
\end{table}

%% file: sections/Behavioral-analysis.tex
\section{Behavior Under Obfuscation and Steering}
\subsection{Model-Generated Reasoning Traces (RQ3.1)}
\label{sec:rq21}

\input{sections/rq21-new}

\begin{figure*}[t]
\centering
\begin{subfigure}[t]{0.499\textwidth}
  \centering
  \includegraphics[width=\linewidth]{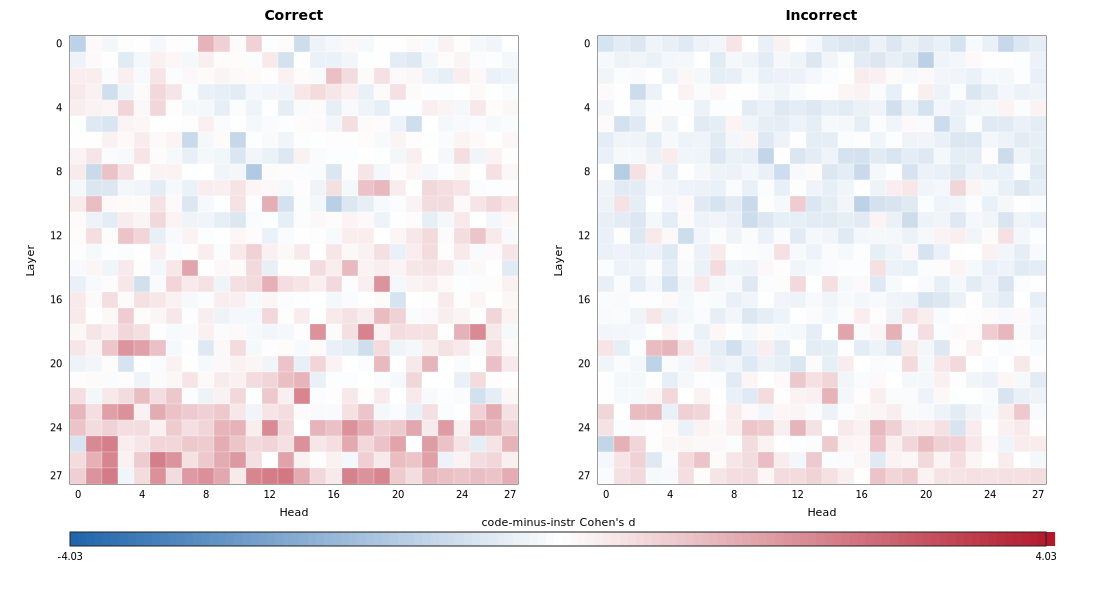}
  \vspace{-21pt}
  \caption{Obfuscated run $VS$ Non-Obfuscated run}
\end{subfigure}\hfill
\begin{subfigure}[t]{0.499\textwidth}
  \centering
  \includegraphics[width=\linewidth]{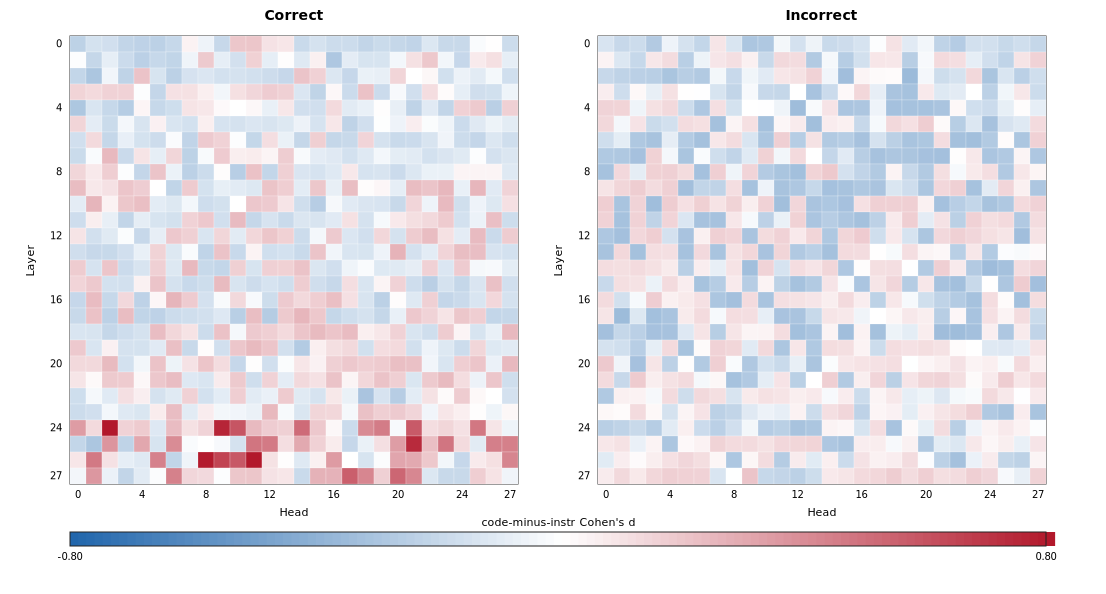}
  \vspace{-21pt}
  \caption{Steered obfuscated run $VS$ Non-Steered Obfuscated run}
\end{subfigure}
\vspace{-12pt}
\caption{Attention-level behavioral changes. Each heatmap reports a head-level code-minus-instruction Cohen's $d$ map for the correct and incorrect sets under one pairwise contrast. Positive values indicate the relatively more code-focus at that head.}
\label{fig:attention}
\end{figure*}

\subsection{Attention Patterns (RQ3.2)}
\label{sec:rq22}




In this experiment, we investigate the model behavioral changes at the attention level in two transitions: (1) from running on the original code to the obfuscated code, and (2) from the obfuscated code to the running on the steered model. Specifically, we analyzed the head-level attention features from \code{Qwen2.5-Coder-7B-Instruct} during our output prediction experiment
under three conditions: original code, obfuscated code, and obfuscated code with steered model. These features are collected as part of the experiment in RQ1, and are stored as a per-head summary tensor for each run.
For each attention head, this tensor records the attention mass allocated to \emph{code tokens}, \emph{instruction tokens}, and \emph{other tokens} in the prompt, as well as other statistics such as attention entropy and head output norm. We then compare these head-level features in the above two transitions. We report all features separately for the sets of \emph{correct} and \emph{incorrect} instances for output prediction.

In particular, we present \emph{Cohen's \(d\) effect size} computed from a per-run \emph{code-bias score}. For a given attention head \(h\) and run \(n\), we define the run-level \emph{code-bias score} as 
\begin{equation}
\mathrm{CodeBias}_{h,n} = \code{attn\_mass\_code}_{h,n} - \code{attn\_mass\_instr}_{h,n},
\end{equation}
where \(\code{attn\_mass\_code}_{h,n}\) is the {\bf \em total attention mass assigned by head \(h\) in run \(n\) to code-specific tokens},~and \(\code{attn\_mass\_instr}_{h,n}\) is the {\bf \emph{total attention mass assigned to natural-language instruction tokens}}. Intuitively, {\em \(\mathrm{CodeBias}_{h,n}\) measures whether that head is more focused on the code itself or on the surrounding instruction text in that run}.
Then, we compute these run-level scores within each group under comparison. Specifically, for groups \(A\) and \(B\) with \(N_A\) and \(N_B\) runs, we define
\[
\mu_{A,h}=\frac{1}{N_A}\sum_{n\in A}\mathrm{CodeBias}_{h,n},
\qquad
\mu_{B,h}=\frac{1}{N_B}\sum_{n\in B}\mathrm{CodeBias}_{h,n},
\]

and let \(\sigma_{A,h}\) and \(\sigma_{B,h}\) denote the corresponding standard deviations of the run-level CodeBias scores within each group. We then compute the pooled standard deviation
\[
s_{p,h}
=
\sqrt{
\frac{
(N_A-1)\sigma_{A,h}^{2}
+
(N_B-1)\sigma_{B,h}^{2}
}{
N_A+N_B-2
}
}
\]
and define the head-level Cohen's \(d\) score as
$
d_h
=
\frac{\mu_{A,h}-\mu_{B,h}}{s_{p,h}}
$.

Cohen's \(d\) is a standardized effect size. Positive values mean that the first-named condition in the comparison is relatively more code-focused at that head, whereas negative values mean that it is relatively more instruction-focused. 




As shown in Figure~\ref{fig:attention}(a), for the correctly predicted cases, the last four layers contain a noticeably higher density of dark red heads. This indicates that, under obfuscation, the model's late layers allocate substantially more attention to code tokens than to other tokens. In contrast, this pattern is largely absent in the incorrect cases. This observation is consistent with our finding in RQ2.1: obfuscation shifts the model toward stronger focus on code elements, encouraging a more statement-by-statement reasoning style analogous to System~2 reasoning in humans.

We next examine the change from the non-steered obfuscated runs to the steered runs in Figure~\ref{fig:attention}(b). For the correct cases, the overall code-focused trend remains, but steering refines this broad shift into a more selective head-level pattern, with some already focused red heads becoming even darker. That is, our steering-guided inference does not simply make all late-layer heads attend more strongly to code. Instead, it concentrates stronger code-focused attention into a smaller subset of late-layer heads, while allowing other heads to remain more sensitive to instruction tokens.

By contrast, the incorrect cases appear more diffuse and less selectively organized. Their pattern is spread more evenly across layers and heads, with much weaker concentration on a specific subset of late-layer heads than in the correct cases. This suggests that successful reasoning is associated not only with stronger attention to code, but also with a selective allocation of that attention.

%% file: sections/rq21-new.tex
While $pass@k$ captures end-task success (as in Section~\ref{sec:rq1}), it does not reveal whether these arise from computationally light, reflexive responses (\emph{System 1}) or from deliberate, structured reasoning (\emph{System 2}). As noted in Section~\ref{sec:intro}, obfuscation is expected to remove surface-level cues and suppress System 1 behavior, whereas steering aims to enforce System 2 reasoning. In this experiment, we examine how these factors influence generated reasoning traces. 


\subsubsection{The Quality of Generated Reasoning Traces} 
Following the restoration trends in Section~\ref{sec:rq1}, we select \code{Qwen2.5-7B} as the representative model for a fine-grained analysis of reasoning behavior. Among the obfuscation strategies, \textit{identifier renaming} induces the largest degradation in performance (Table~\ref{tab:rq3-obf-type}); therefore, to enable a more discriminative analysis, we focus on reasoning traces from this data subset. 
To construct a high-confidence reasoning pool, we only consider output prediction instances for which all three settings (\textit{i.e.}, original, obfuscated, and steered) are answered correctly across all three trials, yielding a candidate subset of {\color{black}{452}} instances across both datasets. Among these, we annotate six observable reasoning traits, grouped into three {\bf positive} behaviors: (P1) \textit{step-by-step execution reasoning}, (P2) \textit{explicit state tracking or verification}, and (P3) \textit{grounding in code statements}; and three {\bf negative} behaviors: (N1) \textit{surface-level lexical shortcuts}, (N2) \textit{ungrounded or hallucinated reasoning steps}, and (N3) \textit{direct answers without justification}. 

To calibrate the annotation rubric, we manually inspect 20 samples and compare them against labels produced by Claude Opus 4.6, observing over 98.5\% agreement; we then apply the same rubric to annotate an additional {\color{black}{150}} samples, which are subsequently verified by human annotators. During labeling, the judge is provided with the original code and corresponding reasoning trace for the original setting, and with the obfuscated code and corresponding reasoning trace for both the obfuscated and steered settings. 

\input{tables/reasoning-quality}

Table~\ref{tab:reasoning-labels} summarizes the resulting label distributions across original programs, obfuscated variants, and obfuscated variants with steering, reporting the fraction of the labeled cases in which each reasoning trait appears in the rationale trace. Overall, {\em steering-guided} generation exhibits the highest prevalence of positive reasoning traits and the lowest incidence of negative ones, aligning with the relative performance trends observed across all program variants in output prediction (Table~\ref{tab:main-eval}). Furthermore, these results support our broader hypothesis: obfuscation can shift the model away from reflexive System~1 lexical matching toward more System~2-style reasoning; however, the plain obfuscated condition still shows more surface-level cues or lexical shortcuts (10.7\%) than either original code (6.3\%) or steered code (2.7\%). This suggests that {\em obfuscation alone does not reliably complete that shift}: when the code becomes confusing, the model can still {\em fall back on shallow priors unless steering keeps attention on semantically relevant statements}. Steering largely recovers performance toward the original model, but it also brings back some hallucination, as reflected by the relatively high rate of ungrounded leap or hallucinated step.

\subsubsection{The Form of Generated Reasoning Traces} To assess the form of reasoning traces, we collect the following behavioral metrics: 
\begin{enumerate}[topsep=0pt, itemsep=0pt,leftmargin=*]
    \item \emph{Reasoning Present} (RP) indicates model engages in System 2--like behavior by producing a substantive explanation ($\geq20$ tokens),
    \item \emph{Direct Answer} (DA) reflects System 1--like behavior, where the model outputs an answer with little or no reasoning,
    \item \emph{Other/Unparsable} (O/U) captures low-quality outputs (e.g., malformed text) that do not link to coherent reasoning or answer,
    \item \emph{\#Tok} represents the average length of the reasoning trace ($\mu\pm\sigma$),
    \item \emph{Avg-Stmt-Mentioned} is the average number of steering-guided statements explicitly mentioned in the reasoning traces,
    \item \emph{Stmt-Recall} is the fraction of steered statements in the traces. 
\end{enumerate}

In this experiment, we use the stored per-run generations from \code{Qwen2.5-7B} under all three evaluation settings.

\input{tables/reasoning-trace-2}

In Table~\ref{tab:rq21-trace-behavior}, for the {\em correctly predicted instances}, we can see a clear shift toward \emph{System 2}--like reasoning, as illustrated by the increasing RP (65.0$\rightarrow$70.8$\rightarrow$76.3) and subsequent decreasing DA (27.7$\rightarrow$21.9$\rightarrow$15.1). In these cases, reasoning length also increases (117$\rightarrow$135$\rightarrow$144), indicating deeper reasoning processes. Moreover, the number of statements in the program slice that are explicitly mentioned in the traces is the highest with steering (on an average, 4.36 per program). {\color{custom-blue}From Table~\ref{tab:rq21-trace-behavior}, the statement recall is 26.2\%. Thus, with 4.36 referenced statements, the model explicitly makes references for every (4.36*100)/26.2 = 16.7 statements on average in the verbalized reasoning trace toward predicting the correct~output}.
The strong alignment between higher RP and improved grounding (\emph{Avg-Stmt-Mentioned} and \emph{Stmt-Recall}) 
suggests {\bf \em successful predictions occur when reasoning is present and is well grounded}.

For \textit{incorrectly predicted instances}, RP drops markedly (by 50.5\%--57.7.\%--46.2\%) alongside a sharp increase in O/U 
and trace length, with reasoning traces also becoming less grounded (lower Avg-Stmt-Mentioned and Stmt-Recall). This implies incorrect predictions are associated with misdirected or incoherent System~2 reasoning. 
Key behavioral differences between correct and incorrect runs are: (1) correct predictions consistently exhibit a higher proportion of RP (65.0--76.3\%) than incorrect ones (43.2--52.2\%), indicating that successful predictions are strongly associated with coherent System 2 reasoning; (2) incorrect runs show a much higher O/U rate, suggesting instability or breakdown in reasoning generation; and (3) incorrect runs tend to have longer yet ineffective rationales. This highlights that {\bf \em reasoning length alone does not guarantee correctness}, and reasoning quality and coherence are critical.

Overall, these findings reveal that {\bf \emph{obfuscation reduces reliance on System~1 behavior, while steering promotes System~2 reasoning}}; but only coherent and well-directed reasoning leads to correct predictions. Accordingly, steering improves both the \emph{frequency} of reasoning (RP) and its \emph{quality} (\textit{i.e.}, grounding), but only for correct predictions. Importantly, this improvement does not arise from simply eliciting more verbose rationales; rather, steering guides the model to focus on code elements most relevant to the task outcome. This underscores that effective reasoning requires not just engaging System~2 thinking, but is grounded in the right elements.

%% file: tables/reasoning-quality.tex
\begin{table}[t]
\centering
\small
\setlength{\tabcolsep}{4pt}
\caption{Pilot reasoning-behavior label rates for original-code reference artifacts, obfuscated plain artifacts, and obfuscated + steering artifacts. Values are percentages.}
\label{tab:reasoning-labels}
\vspace{-9pt}
\resizebox{\columnwidth}{!}{%
\begin{tabular}{lccc}
\toprule
\multirow{2}{*}{Reasoning Traits} & \multicolumn{1}{c}{Orig.} & \multicolumn{1}{c}{Obf. w/o} & \multicolumn{1}{c}{Obf. w/} \\
& & Steering & Steering \\
\midrule
(P1) Step-by-step execution reasoning & 74.2 & 71.3 & 87.3  \\
(P2) Explicit state tracking or verification & 58.9 & 58.7 & 70.7  \\
(P3) Grounded to code statements & 86.5 & 98.0 & 99.3  \\
\midrule
(N1) Surface-cue or lexical shortcut & 6.3 & 10.7 & 2.7  \\
(N2) Ungrounded leap or hallucinated step & 28.7 & 24.0 & 28.7 \\
(N3) Direct answer without reasoning & 16.1 & 2.7 & 1.3 \\
\bottomrule
\end{tabular}%
}
\end{table}

%% file: tables/reasoning-trace-2.tex
\begin{table}[t]
\centering
\setlength{\tabcolsep}{1.5pt}
\caption{{\color{custom-blue}Trace-level behavioral and grounding 
metrics}. "RP"=reasoning present, "DA"=direct answer, "O/U"=other, and "Tok" = avg. rationale tokens among reasoning-present runs (reported as mean$\pm$std). The last two columns measure steered-element grounding among reasoning-present rationales. Percentages are computed within the displayed correctness split. Model=\code{Qwen2.5-Coder-7B}. Corr: Correct prediction.}
\label{tab:rq21-trace-behavior}
\vspace{-6pt}
\scriptsize
\resizebox{\columnwidth}{!}{%
\begin{tabular}{@{}ccc|c|cccccc@{}}
\toprule
Orig. & Obf. & Steer. & Corr. & RP & DA & O/U & \shortstack{\#Tok} & {Avg-Stmt-} & {Stmt-} \\
& & & Pred. & (\%) & (\%) & (\%) &  & Mentioned & Recall \\
\midrule
\cmark & \text{\sffamily --}          &  \text{\sffamily --}  & y & 65.0 & 27.7 & 7.3 & $117 \pm 50$ & 2.737 & 16.5\% \\
-- & \cmark &   \text{\sffamily --}  & y & 70.8 & 21.9 & 7.3 & $135 \pm 74$ & 2.519 & 19.3\% \\
-- & \cmark & \cmark & y & 76.3 & 15.1 & 8.6 & $144 \pm 89$ & {\bf 4.361} & {\bf 26.2}\% \\
\midrule
\cmark &    \text{\sffamily --} &    \text{\sffamily --}    & \text{\sffamily n} & 43.2 & 27.5 & 29.3 & $157 \pm 93$ & 3.091 & 15.1\% \\
-- & \cmark &  \text{\sffamily --}    & \text{\sffamily n} & 44.9 & 24.6 & 30.5 & $179 \pm 102$ & 2.796 & 17.3\% \\
-- & \cmark & \cmark & \text{\sffamily n} & 52.2 & 20.5 & 27.3 & $186 \pm 104$ & 3.279 & 18.9\% \\
\bottomrule

\end{tabular}

}
\end{table}

%% file: sections/RQ4-new.tex
\section{Other Obfuscation Type Generalization (RQ4)}
\label{sec:rq4}

{\color{custom-blue}

\paragraph{Empirical Setting} To evaluate {\tool}'s generalization potential, we chose four new types of obfuscation. First, in the {\bf \em opaque-predicate} obfuscation, a deterministic guard is added to preserve the original execution path while introducing a syntactically plausible decoy branch. For example, a direct statement such as \code{realCode();} can be wrapped as \code{if ((x*x+x)\%2==0) realCode(); else decoy();}. This transformation tests whether the model can ignore an irrelevant but plausible branch without treating the decoy as the semantic target. Second, the {\bf \em loop transformation} obfuscation preserves the iteration order but changes the loop structure. For example, a \code{for} loop is rewritten as a \code{while} loop with separated initialization and update logic.
Third, the {\bf \em branch inversion} obfuscation 
negates the predicate and swaps the branch bodies, e.g., rewriting \code{if (c) A else B} as \code{if (!c) B else A}. This transformation preserves control-flow semantics but requires the model to track predicate polarity correctly. Fourth, the {\bf \em arithmetic/boolean rewriting} replaces direct expressions with equivalent algebraic or bitwise forms. For example, \code{a+b} can be rewritten as \code{(a\string^b)+2*(a\&b)}. This preserves the computed value but forces a model to recognize semantic equivalence.

From each dataset (HumanEval-X and CruxEval-X), we randomly sampled 10 eligible snippets for each of the above obfuscation types, ensuring that each obfuscation type can be applied to the selected code, yielding 80 additional code instances with 320 input-output prediction cases. Each transformed variant is validated against the original oracle so that the evaluated behavior remains unchanged. We used \code{Qwen2.5-Coder-7B-Instruct} as in the experiment in Section~\ref{sec:rq2} and recorded the evaluation metrics for the original, obfuscated, and obfuscated-with-{\tool} settings.

{\em Empirical Result}. As seen in Table~\ref{tab:rq4}, {\tool} improves opaque-predicate cases from 63.75\% to 71.25\% with a restoration ratio of 75\%. This result
shows that the static slice prior does not pre-determine which branch the model should take, yet helps it focus on the output-deciding branches, steering the model from drifting toward the decoy branch. Interestingly, {\tool} makes the model perform even better than itself on the original code (74.58\% vs 73.75\%). Loop transformation spreads the same loop-carried state across different ways of initialization, guard, and update statements. This shows that our static prior does not presolve the loop iteration, yet it helps a model keep its reasoning about statements in the loop structurally connected to the queried output. Table~\ref{tab:rq4} also shows that static priors do not help for arithmetic/boolean rewriting (67.19\% vs. 67.08\%) and branch inversion (71.07\% vs. 70.83\%). These obfuscation types mainly change the semantic meaning of a single expression or predicate. Thus, the static priors across statements do not help.

\begin{table}[t]
\centering
\scriptsize
\setlength{\tabcolsep}{3.5pt}
\caption{{\color{custom-blue}Output prediction effectiveness on additional obfuscation types; values are  P@1 aggregate accuracies (\%); Obf. w/ Steering: steered model on obfuscated code (RQ4)}}
\label{tab:rq4}
\vspace{-9pt}
\resizebox{\columnwidth}{!}{%
\begin{tabular}{lcccc}
\toprule
\multirow{2}{*}{Obfuscation Technique} & \multicolumn{1}{c}{Orig.} & \multicolumn{1}{c}{Obf. w/o} & \multicolumn{1}{c}{Obf. w/} & \multicolumn{1}{c}{Restor.} \\
& & Steering & Steering & Ratio \\
\midrule
Opaque Predicates & \multirow{4}{*}{73.75} & 63.75 & 71.25 & 75.00 \\
Loop Transformation &  & 67.50 & 74.58 & 113.28 \\
Arithmetic/Boolean Rewriting &  & 67.19 & 67.08 & -1.68 \\
Branch Inversion &  & 71.07 & 70.83 & -8.96 \\
\bottomrule
\end{tabular}
}
\end{table}

\begin{figure}[t]
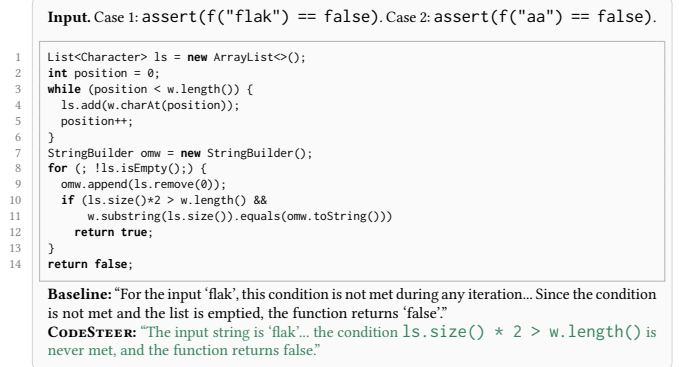

\centering
\scriptsize
\begin{tcolorbox}[
  colback=black!2,
  colframe=black!18,
  boxrule=0.35pt,
  arc=1mm,
  left=1mm,right=1mm,top=0.7mm,bottom=0.7mm,
  fonttitle=\bfseries
]
\textbf{Input.} Case 1: \code{assert(f("flak") == false)}. Case 2: \code{assert(f("aa") == false)}.\\[-2pt]
\begin{lstlisting}[language=Java,
  basicstyle=\ttfamily\tiny,
  keywordstyle=\bfseries,
  columns=fullflexible,
  keepspaces=true,
  breaklines=true,
  showstringspaces=false,
  frame=single,
  framerule=0.25pt,
  rulecolor=\color{black!30},
  aboveskip=0.2em,
  belowskip=0.25em]
List<Character> ls = new ArrayList<>();
int position = 0;
while (position < w.length()) {
  ls.add(w.charAt(position));
  position++;
}
StringBuilder omw = new StringBuilder();
for (; !ls.isEmpty();) {
  omw.append(ls.remove(0));
  if (ls.size()*2 > w.length() &&
      w.substring(ls.size()).equals(omw.toString()))
    return true;
}
return false;
\end{lstlisting}
\textbf{Baseline:} ``For the input `flak', this condition is not met during any iteration... Since the condition is not met and the list is emptied, the function returns `false'.''\\
\textbf{{\tool}:} \textcolor{ForestGreen!70!black}{``The input string is `flak'... the condition \code{ls.size() * 2 > w.length()} is never met, and the function returns false.''}
\end{tcolorbox}
\vspace{-10pt}
\caption{{\color{custom-blue}Loop-transformation reasoning trace example}}
\label{fig:rq4-loop-case}
\end{figure}

{\em A case study}. Figure~\ref{fig:rq4-loop-case} shows a loop-transformation case study (others can be found in project's page~\cite{codesteer2026}). The transformation splits the loop state across a list-building loop and a consuming loop (\code{while} to \code{for} at line 8). As seen in the reasoning traces: without steering, the model correctly follows the helper return value but loses the task mapping from helper return to the assertion predicate, while steering keeps the loop-state reasoning connected to the final assertion being evaluated, leading to correct output prediction.


}


%% file: sections/case-study.tex
\section{Case Study on Vulnerability Analysis}
\label{sec:case-study}

\input{sections/case-study-example}

In this section, we present a case study on how we used {\tool} to perform vulnerability analysis on an obfuscated C program~with buffer overflow. In Figure~\ref{fig:case-study-z9}, the code was obfuscated with alpha-renaming. The local buffer \code{a} has capacity 16, yet the input has 24 characters. We used \code{Qwen3-Coder-30B-A3B-Instruct} in two settings: regular inference and slice-guided steering. {\color{custom-blue}Without steering, the model recognizes the presence of a memory-safety issue, but its reasoning diverges from the concrete execution. Specifically, it overestimates the input length from 24 to 28 characters and retreats to~a generic crash explanation rather than maintaining the precise vulnerability-relevant execution trace in the program. However, via static analysis, the slice prior emphasizes the vulnerability-relevant dataflow around the local buffer \code{a}, the loop index \code{d}, the copy write \code{*(a + d) = e}, the terminating null write \code{*(a + d) = '\textbackslash0'}, and the relevant statements needed to determine if execution reaches these out-of-bound writes. With steering, the model preserves the correct input length and buffer capacity, traces the copy loop to the first out-of-bounds write, and correctly identifies the final NUL terminator write as a second out-of-bounds access}.

%% file: sections/case-study-example.tex
\begin{figure}
\noindent\begin{minipage}{\linewidth}
\captionsetup{type=figure}
\begin{tcolorbox}[
  title={Obfuscated C vulnerability-trace case study},
  colback=black!2,
  colframe=black!15,
  boxrule=0.35pt,
  arc=1mm,
  left=1mm,right=1mm,top=0.7mm,bottom=0.7mm,
  fonttitle=\bfseries,
]
\footnotesize
\textbf{Concrete input:} \code{AAAAAAAAAAAAAAAAAAAAAAAA} (24 characters). 
\end{tcolorbox}
\begin{lstlisting}[language=C,
  basicstyle=\ttfamily\scriptsize,
  keywordstyle=\bfseries,
  columns=fullflexible,
  keepspaces=true,
  breaklines=true,
  showstringspaces=false,
  frame=single,
  framerule=0.35pt,
  rulecolor=\color{black!35},
  xleftmargin=2.5em,
  xrightmargin=0.5em,
  aboveskip=0.35em,
  belowskip=0.35em]
#include <stdio.h>
#define X1 16
#define X2 128
int z9(const char *q) {
    char a[X1];
    int b = 0, c = 0, d = 0;
    for (;;) {
        if (!(q[d] != '\0')) break;
        char e = q[d];
        b += (e >= '0' && e <= '9') ? (e - '0') :
             (e >= 'A' && e <= 'Z') ? 2 :
             ((e ^ e) + 1);

        *(a + d) = e;

        if ((d & 1) && !(d & 1)) {
            b += 999;
        }
        d++;
    }
    *(a + d) = '\0';
    c = ((b ^ 21) < (b ^ 1)) ? 1 : (b > 20);
    int (*p)(const char*, ...) = printf;
    (c ? p : p)(c ? "High score.\n" : "Low score.\n");
    p("Copied string: %s\n", a);
    p("Score: %d\n", b);
    return c;
}

z9((const char *)(y + 0));
    
\end{lstlisting}
\end{minipage}
\begin{tcolorbox}[
  title={Output from the LLM without steering},
  colback=BurntOrange!3,
  colframe=BurntOrange!40,
  boxrule=0.35pt,
  arc=1mm,
  left=1mm,right=1mm,top=0.7mm,bottom=0.7mm,
  fonttitle=\bfseries,
]
\footnotesize
\tracestate{``The input string is treated as having length 28, and the local buffer \code{a} in \code{z9} is treated as having capacity 16. The loop continues copying characters into \code{a[d]} until \code{d} reaches 28, so the input is considered much larger than the destination buffer.}
\tracestate{**Execution trace:**
[30, 31, 32, 33, 5, 6, (7, 8, 9, 10-12, 14, 16, 19) x16, 7, 8, 9, 10-12, 14]}
\tracestate{The program may crash or behave erratically before producing reliable later output.}
\tracestate{Execution stops at the write on `L019` when `d = 16`, because that is treated as the first buffer-overflow point.}

\tracefinal{\textbf{Output:} \texttt{Enter text:} The program may crash or behave erratically before producing reliable later output.''}
\end{tcolorbox}
\begin{tcolorbox}[
  title={Output from the LLM with {\tool}},
  colback=ForestGreen!3,
  colframe=ForestGreen!40,
  boxrule=0.35pt,
  arc=1mm,
  left=1mm,right=1mm,top=0.7mm,bottom=0.7mm,
  fonttitle=\bfseries,
]
\footnotesize
\tracestate{``The \code{z9} function processes the input character by character until the null terminator is encountered. The local buffer \code{a} has size 16, but the concrete input is 24 characters long.}
\tracestate{When the index reaches 16, the next character write goes past the end of \code{a}, and after the loop the terminating NUL write also lands out of bounds.}
\tracestate{The score accumulator still reaches 48, so the score branch is \texttt{High score.}, and the print sink is reached even though the copied string itself is no longer reliable after memory corruption.}
\tracestate{Execution trace:
[30, 31, 32, 33, 5, 6, (7, 8, 9, 10-12, 14, 16, 19) x24, 7, 8, 21, 22, 23, 24, 25, 26, 27, 35]}
\tracefinal{Conclusion: The input length is 24, the destination buffer capacity is 16, the sink is reached, the character copy writes past the end of the buffer, the terminating null byte also writes past the end of the buffer, and the overall result is a buffer overflow.}
\end{tcolorbox}
\vspace{-12pt}
\caption{Buffer-overflow vulnerability analysis}
\label{fig:case-study-z9}
\end{figure}


%% file: sections/related.tex
\vspace{-3pt}
\section{Related Work}
\label{sec:related}

\paragraph{Code Obfuscation and Program Comprehension~\cite{collberg1997taxonomy,collberg2002manufacturing}.}
Nguyen {\em et al.}~\cite{nguyen2026effectcodeobfuscationhuman} reported that obfuscated code reduces humans' accuracy in output prediction and induces a cognitive shift from memory-based reasoning (System 1) to slower, analytical reasoning (System~2). 

\emph{LLMs for Code Understanding.}
LLMs have achieved strong performance across a wide range of code-related tasks~\cite{chen2021evaluatinglargelanguagemodels,roziere2023codegen,li2022codebert,guo2021graphcodebert, LLM4Decompile-Decompiling-Binary-Code-with-Large-Language-Models}. 
We show that LLMs often rely on surface-level patterns, e.g., identifier names and common idioms, rather than deep semantic reasoning. 

\emph{LLMs under Obfuscation.}
Nikiema {\em et al.}~\cite{nikiema2025codebarrierllmsactually} 
showed that performance degrades significantly as obfuscation complexity increases. Similarly, Le {\em et al.}~\cite{le2025namesdisappearrevealingllms} studied how removing meaningful naming and structural cues affects LLM reasoning.
PoorCodeSumEval~\cite{How-Effectively-Do-Code-Language-Models-Understand-Poor-Readability-Code?} evaluates model robustness under degraded code readability, including obfuscation-like transformations. Fang {\em et al.}~\cite{Large-Language-Models-for-Code-Analysis:-Do-LLMs-Really-Do-Their-Job?} examined LLMs' effectiveness in code analysis tasks such as deobfuscation.


\emph{Attention Steering in LLMs.}
In PASTA~\cite{PASTA} and AutoPASTA~\cite{auto-pasta-zhang2024modeltellsattendfaithfulness},  reweighting attention toward selected tokens can improve model faithfulness and reasoning performance. However, these approaches are largely task-agnostic and do not leverage program semantics. 

{\em Static Analysis.} Static analysis aims to estimate program properties over all possible inputs. ~\cite{Static-Analysis-of-Executables-to-Detect-Malicious-Patterns, Augmenting-Decompiler-Output-with-Learned-Variable-Names-and-Types}
In contrast, we position {\tool} as a data-driven \emph{static emulation} that is {\em complementary}. Rather than seeking input-independent guarantees via engineered abstract domains, it predicts the most likely \emph{concrete, per-input} execution. 


%% file: references.bib
@article{collberg1997taxonomy,
  title={A Taxonomy of Obfuscating Transformations},
  author={Collberg, Christian and Thomborson, Clark and Low, Douglas},
  journal={Technical Report},
  year={1997}
}

@inproceedings{collberg2002manufacturing,
author = {Collberg, Christian and Thomborson, Clark and Low, Douglas},
title = {Manufacturing cheap, resilient, and stealthy opaque constructs},
year = {1998},
isbn = {0897919793},
publisher = {Association for Computing Machinery},
address = {New York, NY, USA},
url = {https://doi.org/10.1145/268946.268962},
doi = {10.1145/268946.268962},
booktitle = {Proceedings of the 25th ACM SIGPLAN-SIGACT Symposium on Principles of Programming Languages},
pages = {184--196},
numpages = {13},
location = {San Diego, California, USA},
series = {POPL '98}
}

@inproceedings{
guo2021graphcodebert,
title={GraphCode{\{}BERT{\}}: Pre-training Code Representations with Data Flow},
author={Daya Guo and Shuo Ren and Shuai Lu and Zhangyin Feng and Duyu Tang and Shujie LIU and Long Zhou and Nan Duan and Alexey Svyatkovskiy and Shengyu Fu and Michele Tufano and Shao Kun Deng and Colin Clement and Dawn Drain and Neel Sundaresan and Jian Yin and Daxin Jiang and Ming Zhou},
booktitle={International Conference on Learning Representations},
year={2021},
url={https://openreview.net/forum?id=jLoC4ez43PZ}
}

@misc{roziere2023codegen,
      title={CodeGen: An Open Large Language Model for Code with Multi-Turn Program Synthesis}, 
      author={Erik Nijkamp and Bo Pang and Hiroaki Hayashi and Lifu Tu and Huan Wang and Yingbo Zhou and Silvio Savarese and Caiming Xiong},
      year={2023},
      eprint={2203.13474},
      archivePrefix={arXiv},
      primaryClass={cs.LG},
      url={https://arxiv.org/abs/2203.13474}, 
}

@misc{nikiema2025codebarrierllmsactually,
      title={The Code Barrier: What LLMs Actually Understand?}, 
      author={Serge Lionel Nikiema and Jordan Samhi and Abdoul Kader Kaboré and Jacques Klein and Tegawendé F. Bissyandé},
      year={2025},
      eprint={2504.10557},
      archivePrefix={arXiv},
      primaryClass={cs.SE},
      url={https://arxiv.org/abs/2504.10557}, 
}

@misc{nguyen2026effectcodeobfuscationhuman,
      title={The Effect of Code Obfuscation on Human Program Comprehension}, 
      author={Anh H. N. Nguyen and Jack Le and Ilse Lahnstein Coronado and Tien N. Nguyen},
      year={2026},
      eprint={2603.07668},
      archivePrefix={arXiv},
      primaryClass={cs.SE},
      url={https://arxiv.org/abs/2603.07668}, 
}

@inproceedings{humanevalx,
author = {Zheng, Qinkai and Xia, Xiao and Zou, Xu and Dong, Yuxiao and Wang, Shan and Xue, Yufei and Shen, Lei and Wang, Zihan and Wang, Andi and Li, Yang and Su, Teng and Yang, Zhilin and Tang, Jie},
title = {CodeGeeX: A Pre-Trained Model for Code Generation with Multilingual Benchmarking on HumanEval-X},
year = {2023},
isbn = {9798400701030},
publisher = {Association for Computing Machinery},
address = {New York, NY, USA},
url = {https://doi.org/10.1145/3580305.3599790},
doi = {10.1145/3580305.3599790},
booktitle = {Proceedings of the 29th ACM SIGKDD Conference on Knowledge Discovery and Data Mining},
pages = {5673--5684},
numpages = {12},
location = {Long Beach, CA, USA},
series = {KDD '23}
}

@inproceedings{cruxevalx,
author = {Gu, Alex and Rozi\`{e}re, Baptiste and Leather, Hugh and Solar-Lezama, Armando and Synnaeve, Gabriel and Wang, Sida I.},
title = {CRUXEval: a benchmark for code reasoning, understanding and execution},
year = {2024},
publisher = {JMLR.org},
booktitle = {Proceedings of the 41st International Conference on Machine Learning},
articleno = {659},
numpages = {54},
location = {Vienna, Austria},
series = {ICML'24}
}

@misc{chen2021evaluatinglargelanguagemodels,
      title={Evaluating Large Language Models Trained on Code}, 
      author={Mark Chen and Jerry Tworek and Heewoo Jun and Qiming Yuan and Henrique Ponde de Oliveira Pinto and Jared Kaplan and Harri Edwards and Yuri Burda and Nicholas Joseph and Greg Brockman and Alex Ray and Raul Puri and Gretchen Krueger and Michael Petrov and Heidy Khlaaf and Girish Sastry and Pamela Mishkin and Brooke Chan and Scott Gray and Nick Ryder and Mikhail Pavlov and Alethea Power and Lukasz Kaiser and Mohammad Bavarian and Clemens Winter and Philippe Tillet and Felipe Petroski Such and Dave Cummings and Matthias Plappert and Fotios Chantzis and Elizabeth Barnes and Ariel Herbert-Voss and William Hebgen Guss and Alex Nichol and Alex Paino and Nikolas Tezak and Jie Tang and Igor Babuschkin and Suchir Balaji and Shantanu Jain and William Saunders and Christopher Hesse and Andrew N. Carr and Jan Leike and Josh Achiam and Vedant Misra and Evan Morikawa and Alec Radford and Matthew Knight and Miles Brundage and Mira Murati and Katie Mayer and Peter Welinder and Bob McGrew and Dario Amodei and Sam McCandlish and Ilya Sutskever and Wojciech Zaremba},
      year={2021},
      eprint={2107.03374},
      archivePrefix={arXiv},
      primaryClass={cs.LG},
      url={https://arxiv.org/abs/2107.03374}, 
}

@inproceedings{li2022codebert,
    title = "{C}ode{BERT}: A Pre-Trained Model for Programming and Natural Languages",
    author = "Feng, Zhangyin  and
      Guo, Daya  and
      Tang, Duyu  and
      Duan, Nan  and
      Feng, Xiaocheng  and
      Gong, Ming  and
      Shou, Linjun  and
      Qin, Bing  and
      Liu, Ting  and
      Jiang, Daxin  and
      Zhou, Ming",
    editor = "Cohn, Trevor  and
      He, Yulan  and
      Liu, Yang",
    booktitle = "Findings of the Association for Computational Linguistics: EMNLP 2020",
    month = nov,
    year = "2020",
    address = "Online",
    publisher = "Association for Computational Linguistics",
    url = "https://aclanthology.org/2020.findings-emnlp.139/",
    doi = "10.18653/v1/2020.findings-emnlp.139",
    pages = "1536--1547"
}

@misc{Large-Language-Models-for-Code-Analysis:-Do-LLMs-Really-Do-Their-Job?,
      title={Large Language Models for Code Analysis: Do LLMs Really Do Their Job?}, 
      author={Chongzhou Fang and Ning Miao and Shaurya Srivastav and Jialin Liu and Ruoyu Zhang and Ruijie Fang and Asmita and Ryan Tsang and Najmeh Nazari and Han Wang and Houman Homayoun},
      year={2024},
      eprint={2310.12357},
      archivePrefix={arXiv},
      primaryClass={cs.SE},
      url={https://arxiv.org/abs/2310.12357}, 
}

@ARTICLE{Obfuscation-The-Hidden-Malware,
  author={O'Kane, Philip and Sezer, Sakir and McLaughlin, Kieran},
  journal={IEEE Security and Privacy}, 
  title={Obfuscation: The Hidden Malware}, 
  year={2011},
  volume={9},
  number={5},
  pages={41-47},
  doi={10.1109/MSP.2011.98}}

@inproceedings{An-Observational-Investigation-of-Reverse-Engineers-Process-and-Mental-Models,
author = {Votipka, Daniel and Rabin, Seth and Micinski, Kristopher and Foster, Jeffrey S. and Mazurek, Michelle L.},
title = {An Observational Investigation of Reverse Engineers' Process and Mental Models},
year = {2019},
isbn = {9781450359719},
publisher = {Association for Computing Machinery},
address = {New York, NY, USA},
url = {https://doi.org/10.1145/3290607.3313040},
doi = {10.1145/3290607.3313040},
booktitle = {Extended Abstracts of the 2019 CHI Conference on Human Factors in Computing Systems},
pages = {1--6},
numpages = {6},
location = {Glasgow, Scotland Uk},
series = {CHI EA '19}
}

@inproceedings {Static-Analysis-of-Executables-to-Detect-Malicious-Patterns,
author = {Mihai Christodorescu and Somesh Jha},
title = {Static Analysis of Executables to Detect Malicious Patterns},
booktitle = {12th USENIX Security Symposium (USENIX Security 03)},
year = {2003},
address = {Washington, D.C.},
url = {https://www.usenix.org/conference/12th-usenix-security-symposium/static-analysis-executables-detect-malicious-patterns},
publisher = {USENIX Association},
month = aug
}

@inproceedings {Augmenting-Decompiler-Output-with-Learned-Variable-Names-and-Types,
author = {Qibin Chen and Jeremy Lacomis and Edward J. Schwartz and Claire Le Goues and Graham Neubig and Bogdan Vasilescu},
title = {Augmenting Decompiler Output with Learned Variable Names and Types},
booktitle = {31st USENIX Security Symposium (USENIX Security 22)},
year = {2022},
isbn = {978-1-939133-31-1},
address = {Boston, MA},
pages = {4327--4343},
url = {https://www.usenix.org/conference/usenixsecurity22/presentation/chen-qibin},
publisher = {USENIX Association},
month = aug
}

@inproceedings{LLM4Decompile-Decompiling-Binary-Code-with-Large-Language-Models,
   title={LLM4Decompile: Decompiling Binary Code with Large Language Models},
   url={http://dx.doi.org/10.18653/v1/2024.emnlp-main.203},
   DOI={10.18653/v1/2024.emnlp-main.203},
   booktitle={Proceedings of the 2024 Conference on Empirical Methods in Natural Language Processing},
   publisher={Association for Computational Linguistics},
   author={Tan, Hanzhuo and Luo, Qi and Li, Jing and Zhang, Yuqun},
   year={2024},
   pages={3473--3487} }

@article{Competition-level-code-generation-with-AlphaCode,
   title={Competition-level code generation with AlphaCode},
   volume={378},
   ISSN={1095-9203},
   url={http://dx.doi.org/10.1126/science.abq1158},
   DOI={10.1126/science.abq1158},
   number={6624},
   journal={Science},
   publisher={American Association for the Advancement of Science (AAAS)},
   author={Li, Yujia and Choi, David and Chung, Junyoung and Kushman, Nate and Schrittwieser, Julian and Leblond, Rémi and Eccles, Tom and Keeling, James and Gimeno, Felix and Dal Lago, Agustin and Hubert, Thomas and Choy, Peter and de Masson d'Autume, Cyprien and Babuschkin, Igor and Chen, Xinyun and Huang, Po-Sen and Welbl, Johannes and Gowal, Sven and Cherepanov, Alexey and Molloy, James and Mankowitz, Daniel J. and Sutherland Robson, Esme and Kohli, Pushmeet and de Freitas, Nando and Kavukcuoglu, Koray and Vinyals, Oriol},
   year={2022},
   month=dec, pages={1092--1097} }

@misc{in-context-learning-and-induction-heads,
      title={In-context Learning and Induction Heads}, 
      author={Catherine Olsson and Nelson Elhage and Neel Nanda and Nicholas Joseph and Nova DasSarma and Tom Henighan and Ben Mann and Amanda Askell and Yuntao Bai and Anna Chen and Tom Conerly and Dawn Drain and Deep Ganguli and Zac Hatfield-Dodds and Danny Hernandez and Scott Johnston and Andy Jones and Jackson Kernion and Liane Lovitt and Kamal Ndousse and Dario Amodei and Tom Brown and Jack Clark and Jared Kaplan and Sam McCandlish and Chris Olah},
      year={2022},
      eprint={2209.11895},
      archivePrefix={arXiv},
      primaryClass={cs.LG},
      url={https://arxiv.org/abs/2209.11895}, 
}

@misc{Interpretability-in-the-Wild:-A-Circuit-for-Indirect-Object-Identification-in-GPT-2-Small,
      title={Interpretability in the Wild: a Circuit for Indirect Object Identification in GPT-2 small}, 
      author={Kevin Wang and Alexandre Variengien and Arthur Conmy and Buck Shlegeris and Jacob Steinhardt},
      year={2022},
      eprint={2211.00593},
      archivePrefix={arXiv},
      primaryClass={cs.LG},
      url={https://arxiv.org/abs/2211.00593}, 
}

@misc{Towards-Automated-Circuit-Discovery-for-Mechanistic-Interpretability,
      title={Towards Automated Circuit Discovery for Mechanistic Interpretability}, 
      author={Arthur Conmy and Augustine N. Mavor-Parker and Aengus Lynch and Stefan Heimersheim and Adrià Garriga-Alonso},
      year={2023},
      eprint={2304.14997},
      archivePrefix={arXiv},
      primaryClass={cs.LG},
      url={https://arxiv.org/abs/2304.14997}, 
}

@inproceedings{control-flow-flattening,
author = {László, Tímea and Kiss, Ákos},
year = {2007},
month = {06},
pages = {},
title = {Obfuscating C++ Programs via Control Flow Flattening},
volume = {30},
journal = {Annales Universitatis Scientiarum Budapestinensis de Rolando Eötvös Nominatae. Sectio Computatorica}
}

@misc{PASTA,
      title={Tell Your Model Where to Attend: Post-hoc Attention Steering for LLMs}, 
      author={Qingru Zhang and Chandan Singh and Liyuan Liu and Xiaodong Liu and Bin Yu and Jianfeng Gao and Tuo Zhao},
      year={2024},
      eprint={2311.02262},
      archivePrefix={arXiv},
      primaryClass={cs.CL},
      url={https://arxiv.org/abs/2311.02262}, 
}

@misc{auto-pasta-zhang2024modeltellsattendfaithfulness,
      title={Model Tells Itself Where to Attend: Faithfulness Meets Automatic Attention Steering}, 
      author={Qingru Zhang and Xiaodong Yu and Chandan Singh and Xiaodong Liu and Liyuan Liu and Jianfeng Gao and Tuo Zhao and Dan Roth and Hao Cheng},
      year={2024},
      eprint={2409.10790},
      archivePrefix={arXiv},
      primaryClass={cs.CL},
      url={https://arxiv.org/abs/2409.10790}, 
}

@inproceedings{How-Effectively-Do-Code-Language-Models-Understand-Poor-Readability-Code?,
author = {Hu, Chao and Chai, Yitian and Zhou, Hao and Meng, Fandong and Zhou, Jie and Gu, Xiaodong},
title = {How Effectively Do Code Language Models Understand Poor-Readability Code?},
year = {2024},
isbn = {9798400712487},
publisher = {Association for Computing Machinery},
address = {New York, NY, USA},
url = {https://doi.org/10.1145/3691620.3695072},
doi = {10.1145/3691620.3695072},
booktitle = {Proceedings of the 39th IEEE/ACM International Conference on Automated Software Engineering},
pages = {795--806},
numpages = {12},
location = {Sacramento, CA, USA},
series = {ASE '24}
}

@misc{le2025namesdisappearrevealingllms,
      title={When Names Disappear: Revealing What LLMs Actually Understand About Code}, 
      author={Cuong Chi Le and Minh V. T. Pham and Cuong Duc Van and Hoang N. Phan and Huy N. Phan and Tien N. Nguyen},
      year={2025},
      eprint={2510.03178},
      archivePrefix={arXiv},
      primaryClass={cs.SE},
      url={https://arxiv.org/abs/2510.03178}, 
}

@misc{codesteer2026,
  author       = {Anonymous},
  title        = {{CodeSteer}},
  year         = {2026},
  month        = mar,
  publisher    = {Zenodo},
  doi          = {10.5281/zenodo.19340781},
  url          = {https://doi.org/10.5281/zenodo.19340781}
}

@article{li2022competition,
   title={Competition-level code generation with AlphaCode},
   volume={378},
   ISSN={1095-9203},
   url={http://dx.doi.org/10.1126/science.abq1158},
   DOI={10.1126/science.abq1158},
   number={6624},
   journal={Science},
   publisher={American Association for the Advancement of Science (AAAS)},
   author={Li, Yujia and Choi, David and Chung, Junyoung and Kushman, Nate and Schrittwieser, Julian and Leblond, Rémi and Eccles, Tom and Keeling, James and Gimeno, Felix and Dal Lago, Agustin and Hubert, Thomas and Choy, Peter and de Masson d'Autume, Cyprien and Babuschkin, Igor and Chen, Xinyun and Huang, Po-Sen and Welbl, Johannes and Gowal, Sven and Cherepanov, Alexey and Molloy, James and Mankowitz, Daniel J. and Sutherland Robson, Esme and Kohli, Pushmeet and de Freitas, Nando and Kavukcuoglu, Koray and Vinyals, Oriol},
   year={2022},
   month=Dec, pages={1092--1097} }
